\documentclass[twocolumn,amsmath,amssymb,preprintnumbers]{revtex4}
\usepackage{graphicx}
\usepackage{datetime}
\usepackage{hyperref,breakurl}
\graphicspath{{./figs_pdf/}}

\begin{document}
\title{Theoretical description of heavy impurity transport and its application to the modelling of tungsten in JET and ASDEX Upgrade}

\author{F.J.~Casson$^{1,2}$, C.~Angioni$^2$, E.A.~Belli$^3$, R.~Bilato$^2$, P.~Mantica$^4$, T.~Odstrcil$^2$, T.~P{\"u}tterich$^2$, 
M.~Valisa$^5$, L.~Garzotti$^1$, C.~Giroud$^1$, J.~Hobirk$^2$, C.F.~Maggi$^2$, J.~Mlynar$^6$, M.L.~Reinke$^7$,
JET EFDA contributors\footnote{See Appendix of F. Romanelli et al., Proc. 24th IAEA FEC, San Diego, US, 2012} 
and ASDEX-Upgrade team\footnote{See A. Kallenbach et al., Proc. 24th IAEA FEC, San Diego, US, 2012}
}
\affiliation{JET-EFDA Culham Science centre, Abingdon; UK}
\affiliation{$^1$ CCFE, Culham Science Centre, Abingdon, Oxon, OX14 3DB, UK}
\affiliation{$^2$ Max-Planck-Institut f\"{u}r Plasmaphysik, Garching, Germany}
\affiliation{$^3$ General Atomics, PO Box 85608, San Diego, CA 92186-5608, USA}
\affiliation{$^4$ Istituto di Fisica del Plasma, CNR/ENEA, Milano, Italy}
\affiliation{$^5$ Consorzio RFX-CNR/ENEA, I-35127 Padova, Italy}
\affiliation{$^6$ IPP.CR, Inst. of Plasma Physics AS CR, Prague, Czech Republic}
\affiliation{$^7$ University of York, Department of Physics, Heslington, York, YO10 5DD, U.K.}

\begin{abstract}

Recent developments in theory-based modelling of core heavy impurity transport are presented, and shown to be necessary for quantitative description of present experiments in JET and \mbox{ASDEX} Upgrade.  The treatment of heavy impurities is complicated by their large mass and charge, which result in a strong response to plasma rotation or any small background electrostatic field in the plasma, such as that generated by anisotropic external heating. These forces lead to strong poloidal asymmetries of impurity density, which have recently been added to numerical tools describing both neoclassical and turbulent transport.  
Modelling predictions of the steady-state two-dimensional tungsten impurity distribution are compared with experimental densities interpreted from soft X-ray diagnostics.  The modelling identifies neoclassical transport enhanced by poloidal asymmetries as the dominant mechanism responsible for tungsten accumulation in the central core of the plasma.  Depending on the bulk plasma profiles, neoclassical temperature screening can prevent accumulation, and can be enhanced by externally heated species, demonstrated here in ICRH plasmas.

\end{abstract}

\maketitle

\section{Introduction}

Tungsten (W) has good properties as a plasma facing component due to its high heat tolerance, 
low erosion rate, and low hydrogen retention. Tungsten will be used in ITER, is a candidate material for a
fusion reactor, and is presently used in the ASDEX Upgrade (AUG) tokamak and the recently installed ITER-like wall (ILW) at JET. Since tungsten and other high-Z ions radiate strongly, their concentration in
a fusion plasma must be minimised, and central accumulation must be avoided to ensure
stable operation and good performance. For ITER scenario planning, it is
therefore vital to have an understanding of impurity transport underpinned by
comprehensive theoretical models \cite{shimada_iter_2009}. As a prerequisite for reliable predictions, it is important
that these models be quantitatively validated against existing experiments. 

Due to their large mass and charge, heavy impurities such as W experience strong inertial and electrostatic forces, with the result that their densities are not flux functions, but have strong poloidal asymmetries.  
In a rotating plasma, the centrifugal force (CF) is well known since Refs. \cite{hinton_neoclassical_1985,wesson_poloidal_1997} to cause impurity localisation on the low field side (LFS).  
The associated increase in neoclassical transport has long been worked out in analytic models, 
\cite{chang_enhancement_1983,hinton_neoclassical_1985,helander_neoclassical_1998,helander_bifurcated_1998, 
romanelli_effects_1998,fulop_nonlinear_1999,fulop_effect_2002}
but has not usually been included in the numerical tools used for scenario modelling and validation studies
\cite{parisot_experimental_2008,valisa_metal_2011}. 
More recently, temperature anisotropies in a minority species
heated by Ion Cyclotron Resonance Heating (ICRH) have been observed to create a poloidal electric field leading to high field side (HFS) localisation of heavy impurities \cite{ingesson_comparison_2000,reinke_poloidal_2012}.
The theory of ICRH induced anisotropy has since been clarified \cite{bilato_modelling_2014} and impurity transport theories have been extended to account for these effects \cite{fulop_effect_2011,angioni_analytic_2012,mollen_effect_2012,angioni_neoclassical_2014,belli_ps_2014,mollen_impurity_2014}.

For light impurities, where turbulence dominates neoclassical transport, model validation is progressing well \cite{guirlet_anomalous_2009,angioni_gyrokinetic_2011,howard_quantitative_2012,casson_validation_2013,henderson_neoclassical_2014}.  
Meanwhile, results from the JET-ILW have renewed interest in heavy impurity transport, and now motivated the application \cite{angioni_tungsten_2014} of the transport codes {\sc gkw} \cite{peeters_nonlinear_2009} and {\sc neo} \cite{belli_kinetic_2008,belli_full_2012} which both include comprehensive treatments of poloidal asymmetries \cite{casson_gyrokinetic_2010,belli_eulerian_2009}.  

The first validation of the {\sc gkw} + {\sc neo} model for heavy impurities
was made in Ref. \cite{angioni_tungsten_2014}, in which the model quantitatively explained the evolution of core W in the JET hybrid H-mode (NBI heating only).  There, neoclassical transport enhanced by CF effects was shown
to be the primary cause of W accumulation (defined here as strongly peaked W profiles in the \textit{central core}), and the need to include poloidal asymmetries in the impurity transport models was demonstrated.  

In this work, {\sc gkw} + {\sc neo} model validation is extended
by application to the improved H-mode scenario with current overshoot in AUG (Sec. IV), and the ICRH heated baseline H-mode in JET (Sec. V).
New minority heating effects are included in the model for the JET cases, where
central ICRH heating can prevent central W accumulation \cite{putterich_observations_2013,giroud_this_2014,goniche_this_2014}, and can reverse the sign of impurity convection \cite{puiatti_analysis_2006,valisa_metal_2011}.
Predicted two-dimensional impurity density distributions are compared with tomography from soft X-ray diagnostics.
Sec. II outlines the effects of poloidal asymmetries on neoclassical transport, Sec. III describes the modelling setup, and new results are presented in Secs. IV (AUG) and V (JET).

\section{Impact of poloidal asymmetries on neoclassical transport \label{sec.theory}}

In this section, we focus on the significant effects of the poloidal asymmetries on neoclassical transport.  The asymmetry effects on turbulent transport are also included in our {\sc gkw} modelling, but their impact on turbulence is less dramatic (see Fig. \ref{fig.aug_transport}), and can go in both directions, due to subtle interactions between kinetic profiles and magnetic field shear \cite{fulop_effect_2011,angioni_analytic_2012,mollen_effect_2012,mollen_impurity_2014}. 

Neoclassical transport is a flux surface average of local flux vectors which reverse sign from HFS to LFS, so changes in the poloidal density distribution re-weight this average, changing both the sign and magnitude of the net flux \cite{chang_enhancement_1983,hinton_neoclassical_1985,helander_neoclassical_1998,helander_bifurcated_1998, 
romanelli_effects_1998,fulop_nonlinear_1999,fulop_effect_2002}. 
We use the model for poloidal asymmetries, presented in Ref. \cite{bilato_modelling_2014};
solving the parallel force balance, an anisotropically heated species approximated by a bi-Maxwellian (with $T_\parallel$, $T_\perp$)
has poloidally varying equilibrium density
\begin{multline}
 n(\theta) = 
n_{R0} \frac{T_{\perp}(\theta)}{T_{\perp R0}} \cdot \\
 \exp\left(-\frac{e Z \Phi(\theta)}{T_\parallel} + \frac{m \Omega^2 (R(\theta)^2-R_0^2)}{2 T_\parallel}\right)
\label{eq.bilato}
\end{multline}
where $\theta$ is poloidal angle, $\Omega$ is plasma angular rotation frequency, $R$ is major radius, $R_0$ represents LFS values, and
\begin{equation}
 \frac{T_{\perp}(\theta)}{T_{\perp R0}} = \left [ \frac{T_{\perp
 R0}}{T_{\parallel}} + \left(1 - \frac{T_{\perp R0}}{T_{\parallel}} \right )\frac{B_{R0}}{B(\theta)} \right ]^{-1}.
\end{equation}
A minority species with $T_\perp > T_\parallel$ is localized on the LFS and creates a poloidally varying potential $\Phi$
which pushes high Z impurities towards the HFS (if stronger than the centrifugal force).  
Eq. \ref{eq.bilato} is also valid for all isotropic species, which have $T_\perp / T_\parallel = T_{\perp}(\theta)/T_{\perp R0} = 1$.  Both {\sc gkw} and {\sc neo} solve for $\Phi$ for an arbitrary number of species using a quasi-neutral root-finding algorithm \cite{casson_turbulent_2011}.

\begin{figure}[tb] 
\begin{center}
\includegraphics[width=9.5truecm,trim=32 0 15 0,clip=true]{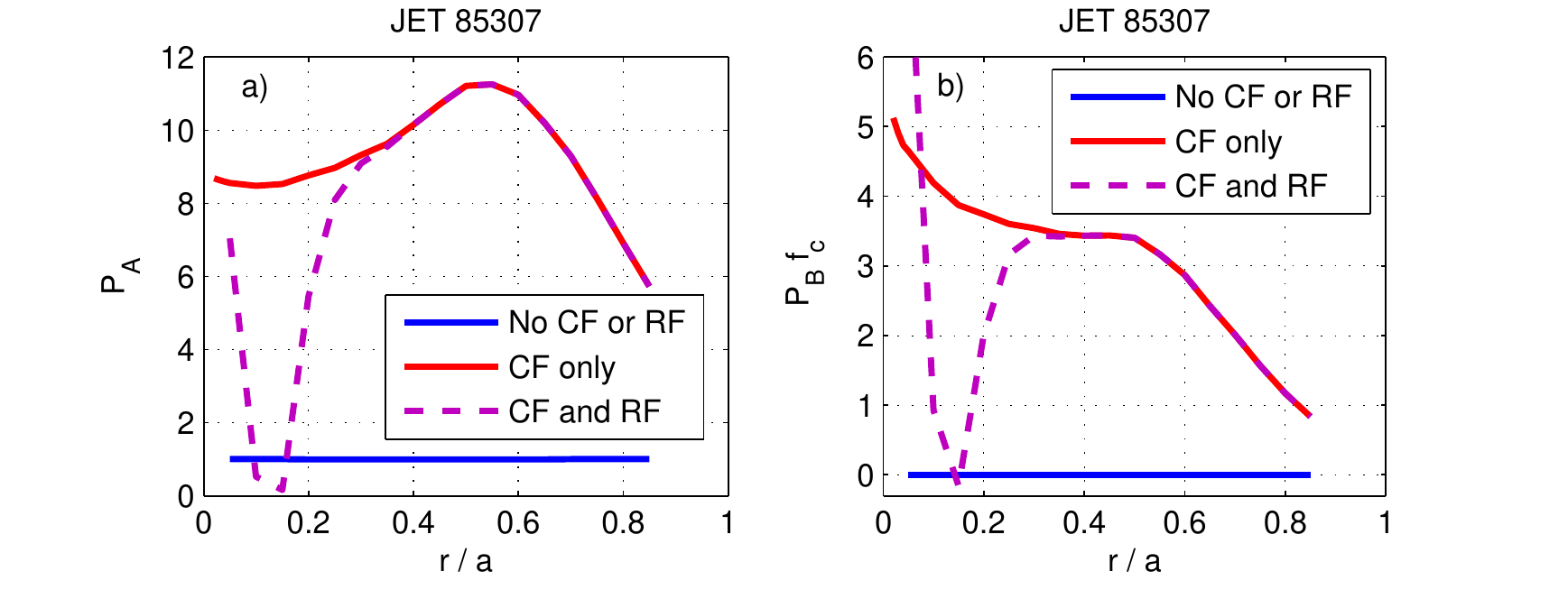}
\caption{Poloidal asymmetry geometrical factors $P_A$ and $P_B f_c$ for neoclassical transport calculated by {\sc gkw} for the JET case with central ICRH in Sec. \ref{sec.jet}.  Poloidal asymmetries can be generated by rotation (CF) or minority heating (RF).} 
\label{fig.pol_factors}
\end{center}
\end{figure} 

Neoclassical impurity transport theory has recently been updated to elaborate the case of HFS impurity localisation \cite{angioni_neoclassical_2014}:  When trace impurities are in the deep Pfirsch-Schl\"{u}ter (PS) regime, and Deuterium is in the Banana regime, the neoclassical impurity transport (with a simplified collision model valid at large aspect ratio) can be summarized as \cite{angioni_neoclassical_2014}
\begin{multline}
R \langle {\bf \Gamma}_z^{\rm neo} \cdot \nabla r \rangle
\propto 
n_i T_i \nu_{ii} Z \Biggl [
P_A \left( -\frac{R}{L_{n_i}} + \frac{1}{2} \frac{R}{L_{T_i}} + \frac{1}{Z} \frac{R}{L_{n_z}} \right) \\
- 0.33 P_B f_{c} \frac{R}{L_{T_i}} \Biggr ]
\label{eq.nc_transport}
\end{multline}
where $f_{c}$ is the circulating (non-trapped) fraction, and $P_A$, $P_B$ are geometrical factors related to the poloidal asymmetry  

\begin{equation}
2 P_A \epsilon^2 = \left \langle \frac{n_z}{B^2} \right \rangle
\frac{\langle B^2 \rangle}{\langle n_z \rangle}   
-
\left[ \left \langle \frac{B^2}{n_z} \right \rangle
\frac{\langle n_z \rangle}{\langle B^2 \rangle}  \right]^{-1},
\end{equation}
\begin{equation}
2 P_B \epsilon^2 =  1 - 
\left[ \left \langle \frac{B^2}{n_z} \right \rangle
\frac{\langle n_z \rangle}{\langle B^2 \rangle} \right]^{-1}.
\end{equation}

For clarity, we have here re-introduced the diffusive term which is ordered small at large $Z$ (and was dropped in Ref. \cite{angioni_neoclassical_2014}).  The usual neoclassical pinch, temperature screening and diffusion (respectively) then appear multiplied by the factor $P_A$.  In addition, a term 
$\propto P_B$ is present, which reduces the temperature screening,
with the coefficient 0.33 applying in the trace limit with D in the Banana regime.
For the poloidally symetric case, $P_A = 1$, $P_B = 0$, and standard neoclassical impurity transport is recovered.

\begin{figure}[tb] 
\begin{center}
\includegraphics[width=7truecm]{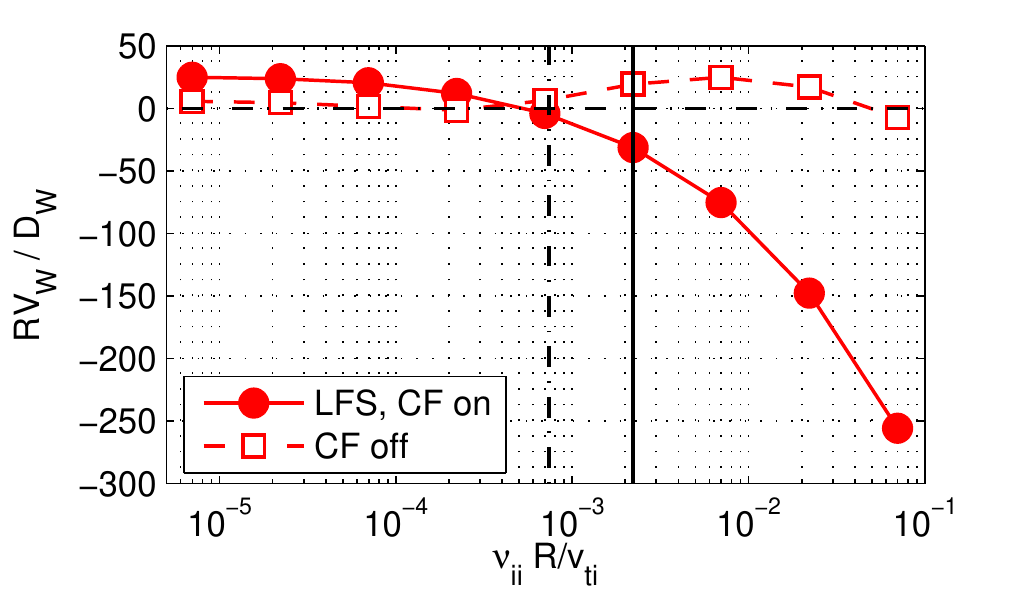}
\caption{Collisionality scan of the {\sc neo}-only peaking factor ($R/L_{n_W} = - RV_W/D_W$) at mid-radius for the JET hybrid case presented in Ref. \cite{angioni_tungsten_2014}, both with and without centrifugal effects.  The vertical lines indicate the collisionality in hybrid (dashed) and baseline (solid) H-modes.} 
\label{fig.dv_theory}
\end{center}
\end{figure} 

In  Ref. \cite{angioni_neoclassical_2014}, the asymmetry factors $P_A$, $P_B$, were calculated for a circular plasma in the limits of weak and strong poloidal asymmetries.  Here, we present the values in full geometry, with realistic anisotropy calculated by {\sc gkw} (Fig. \ref{fig.pol_factors}) for the JET NBI + ICRH case in Sec. \ref{sec.jet}.  From $P_A$ (Fig. \ref{fig.pol_factors}a), it is evident that CF effects greatly increase the neoclassical pinch and diffusion; from $P_B$ (Fig. \ref{fig.pol_factors}b) it is clear that the neoclassical $V/D$ ratio can also be changed, since the extra $f_c P_B$ term (largest at small $r/a$) reduces the effective temperature screening \textit{relative to the other terms} (Fig. \ref{fig.dv_theory}):  At high collisionality, with W in the deep PS regime, Ref. \cite{angioni_neoclassical_2014} applies and the effective temperature screening is reduced by CF effects, making the convection more inward.   At lower collisionality, as the impurities move out of the PS regime, Ref. \cite{angioni_neoclassical_2014} no longer applies, and the numerical {\sc neo} results 
show that 
the CF effects can reverse sign and reduce the neoclassical $R/L_{n_W} = - RV_W/D_W$ (which might be beneficial in a hotter reactor). 
For JET H-modes, typical collisionalities are marked in Fig. \ref{fig.pol_factors}, and indicate that the JET hybrid scenario in Ref. \cite{angioni_neoclassical_2014} is close to a crossover where $R/L_{n_W}$ is not significantly affected by the CF effects (although both $V$ and $D$ are increased by an order of magnitude).  For the AUG improved H-mode in Sec. \ref{sec.jet}, the collisionality is similar to the JET hybrid, but
the parameters differ such that the CF effects decrease $R/L_{n_W}$.  For the JET baseline H-mode (as in Sec. \ref{sec.jet} and \cite{mantica_this_2014}), the CF effects ($P_B$ term) reduce temperature screening and increase $R/L_{n_W}$, with a stronger effect at smaller minor radius.  Given this collisionality and parameter dependence, it is clear that there is no simple scaling fix for less sophisticated neoclassical models that exclude CF effects, and that poloidal asymmetries cannot be neglected in calculations of heavy impurity transport.

\section{Modelling Methodology \label{sec.method}}

We model steady-state H-mode plasmas using gyrokinetic and neoclassical models including both the rotation-induced and anistropy-induced poloidal asymmetries discussed above.   The turbulent transport is computed with the gyrokinetic code {\sc gkw} \cite{peeters_nonlinear_2009} including all rotational effects 
\cite{peeters_influence_2009, camenen_impact_2009, casson_gyrokinetic_2010, angioni_analytic_2012}, here run in its local, quasilinear (6 modes), and electrostatic limits.
The neoclassical transport is computed with the local drift kinetic code {\sc neo} \cite{belli_kinetic_2008,belli_eulerian_2009,belli_full_2012}.
In both codes, ions, electrons and impurities are all modelled kinetically, with W in the trace limit. At each radial location, the W impurity is
modelled in a single average charge state $Z_W$ between 24 (edge) and $46$ (core) of the coronal equilibrium (the charge state range is narrow $\Delta Z < 5$ at the relevant $T_e$). In GKW, $Z_{\rm eff}$ is used only in the collision operator.  For {\sc neo}, an additional species Be (for JET) or B (for AUG) is included to match the measured $Z_{\rm eff}$ profile.  For the JET cases, the hydrogen minority is also present in all simulations at concentrations determined from the isotope shift in the edge Balmer-$\alpha$ spectroscopy.  

The trace limit allows linearisation of the W transport and is appropriate for most conditions, since W concentrations are usually small ($n_W/n_e < 10^{-4}$), except at the end of extreme accumulation phases \cite{putterich_observations_2013}.  The impurity transport is then linearly 
decomposed into convective and diffusive components 
\begin{equation}
R \frac{\Gamma_Z}{n_Z} = D_Z^{\rm GKW} \frac{R}{L_{n_Z,R0}}  + D_Z^{\rm NEO} \frac{R}{L_{n_Z,R0}}  + RV_Z^{\rm  GKW} + RV_Z^{\rm  NEO}
\label{eq.dv}
\end{equation}
which are extracted from the two codes using the fluxes of trace species with different gradients.  For a poloidally asymmetric distribution, ${R}/{L_{n_Z}}$ depends on $\theta$; in Eq. \ref{eq.dv} we use the value defined \textit{at the LFS} (most convenient for the codes).  This choice also defines $D$ and $V$; for transport codes which use flux surface averaged densities, post-processing transformations for $D$ and $V$ are required (defined in Ref. \cite{angioni_tungsten_2014}).  The kinetic profiles and rotation of the bulk plasma (and minority, in Sec. \ref{sec.jet}) are modelling inputs, 
and the four transport coefficients in Eq. \ref{eq.dv} are outputs.  
The modelling then combines turbulent and neoclassical transport channels using the anomalous heat diffusivity $\chi_i^{\rm an}$ from an interpretive power balance calculation (here using {\sc jetto} \cite{cenacchi_jetto:_1988,romanelli_jintrac:_2014} or {\sc astra} \cite{fable_progress_2012}) to normalize the two transport channels relative to each other \cite{angioni_gyrokinetic_2011, casson_validation_2013, angioni_tungsten_2014}.
The ratio of combined convection to combined diffusion is a prediction of the steady-state impurity logarithmic density gradient at the low field side
\begin{equation}
 \frac{R}{L_{n_Z}} =
- \frac{
 \frac{\chi_{\rm i \, an}}{\chi_{\rm i \, NEO} }\cdot
 \frac{RV_{\rm Z \, GKW}}{\chi_{\rm i \, GKW}} +
 \frac{RV_{\rm Z \, NEO}}{\chi_{\rm i \, NEO}}
}
{
 \frac{\chi_{\rm i \, an}}{\chi_{\rm i \, NEO} }\cdot
 \frac{D_{\rm Z \, GKW}}{\chi_{\rm i \, GKW}} +
 \frac{D_{\rm Z \, NEO}}{\chi_{\rm i \, NEO}}
}.
\end{equation}
The modelling is performed at up to 20 radial locations from $r/a = 0.02$ to $ r/a = 0.85$.  Given a boundary value, the LFS density gradient is integrated across the profile to predict a LFS impurity profile.  Finally, the poloidal variation is integrated using the outputs of the quasi-neutrality solver and Eq. \ref{eq.bilato}, to produce a 2D prediction of the impurity distribution. 
For comparison to soft X-ray (SXR) measurements, the SXR 
emission is forward modelled by a simple multiplication with 
a $T_e$-dependent cooling factor and the $n_e$ profile.

To finish this section, we offer some general comments on the modelling sensitivities.    
An example sensitivity test is shown in Fig. \ref{fig.jet_cf_compare}, but we do not have space to present detailed sensitivity studies here.
The key sensitivities are to the logarithmic gradient inputs of bulk ion density $n_i \propto n_e$ and temperature $T_i$, which determine both turbulent stability and neoclassical transport.  In the method described above, the usual sensitivity of turbulence to gradients is removed by the power balance normalisation, but the radial location of the turbulence stability boundary can be moved by $\sim \pm 0.1 r/a$ by changes in the gradients.  Once unstable, the quasilinear turbulent transport ratios are robust to small changes in inputs.  It is the neoclassical transport that is responsible for the bigger uncertainties in the W predictions.

In our experience, the central region of the plasma $r/a < 0.3$ is particularly challenging for quantitative validation for a combination of reasons:  In this region, where turbulence is usually absent, the delicate balance between density and temperature gradients makes neoclassical transport very sensitive to input profiles.  Kinetic measurements in the deep core (vital as inputs for these simulations) are often unavailable or inaccurate, and the profile fits are particularly sensitive to the choice of boundary conditions and the location of the magnetic axis in the equilibrium reconstruction.  The steady-state required for simple profile prediction cannot be reached in the presence of sawteeth.  The validity of the neoclassical model close to the axis (often questioned) is a relatively minor problem by contrast: in the JET cases presented here the size of the potato orbit region is around 1cm for D, and 0.4cm for W.

\section{W Transport under NBI heating, ASDEX Upgrade improved H-mode \label{sec.aug}}

\begin{figure}[tbh] 
\begin{center}
\includegraphics[width=6truecm,trim=0 110 0 0,clip=true]{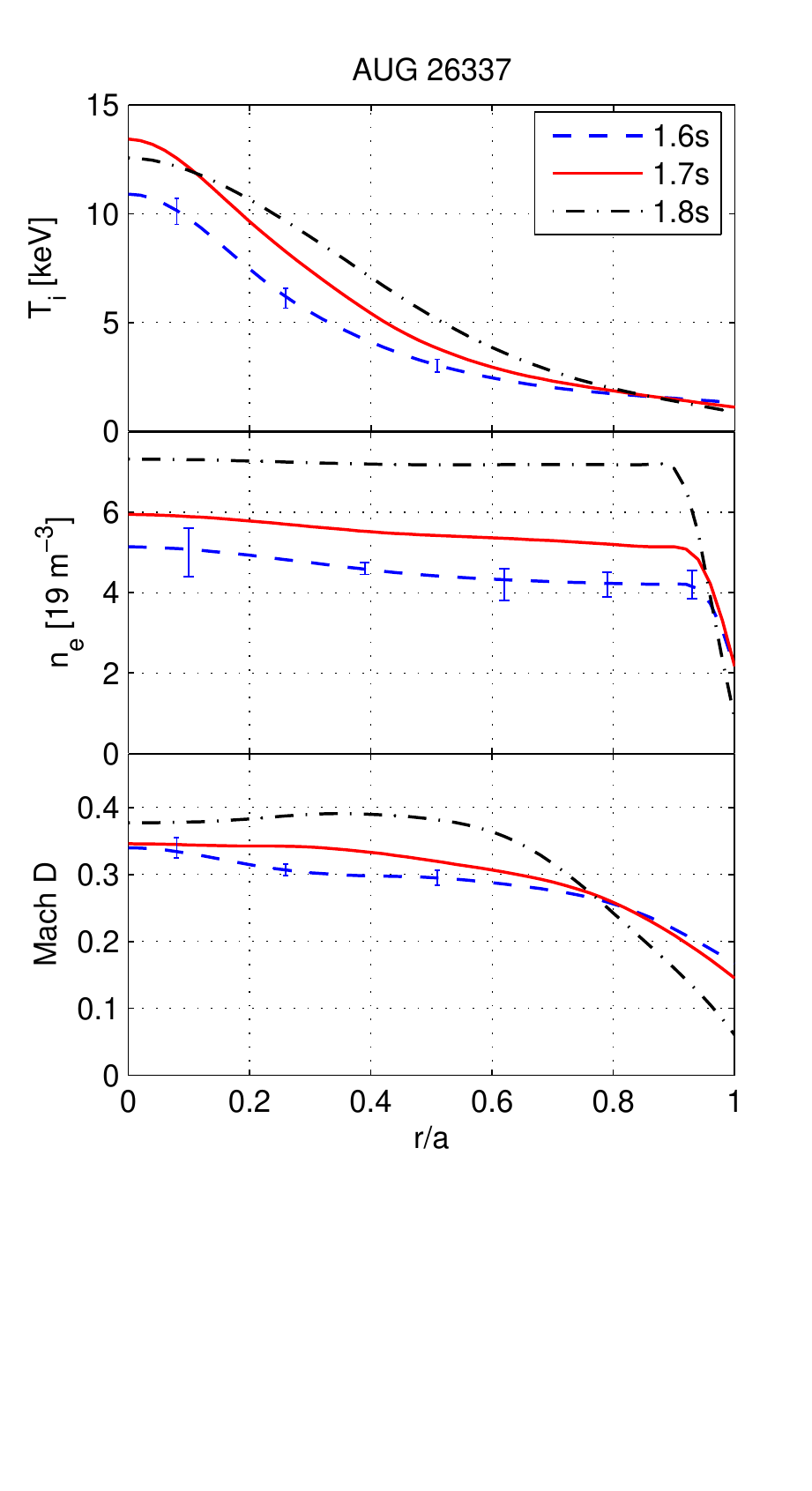}
\caption{Input profiles for simulated timeslices in AUG 26337, with indicitive error bars for selected points.} 
\label{fig.aug_profs}
\end{center}
\end{figure}  

In this section we present modelling of the AUG improved H-mode discharge 26337 
presented in Ref. \cite{hobirk_overview_2012}.  In these discharges, the ``current overshoot'' ramp-up technique is used
to produce a very flat central q-profile $\sim 1$ and a transient period of improving confinement.  
Tungsten is not observed to accumulate, suggested in Ref. \cite{hobirk_overview_2012} to be due to 
the enhancement of neoclassical transport due to the rotation. 
To examine this hypothesis, we model three time slices at the start of the current flattop (ELM-free H-mode), during which the confinement is improving as the NBI power is stepped up (t=1.6s: 5MW; t=1.7s, 7.5MW, t=1.8s, 10MW).  The density profile is quite flat 
but the temperature profile is increasingly peaked (Fig. \ref{fig.aug_profs}).  The low densities and high NBI power (much larger than the 800kW central ECRH) result in large plasma rotation, with some of the highest thermal Mach numbers ($M_D = \Omega R / \sqrt{2 T_D / m_D}$) for AUG, reaching 0.3-0.4 in the core.

\begin{figure}[tbh] 
\begin{center}
\includegraphics[width=8truecm,trim=0 10 0 0,clip=true]{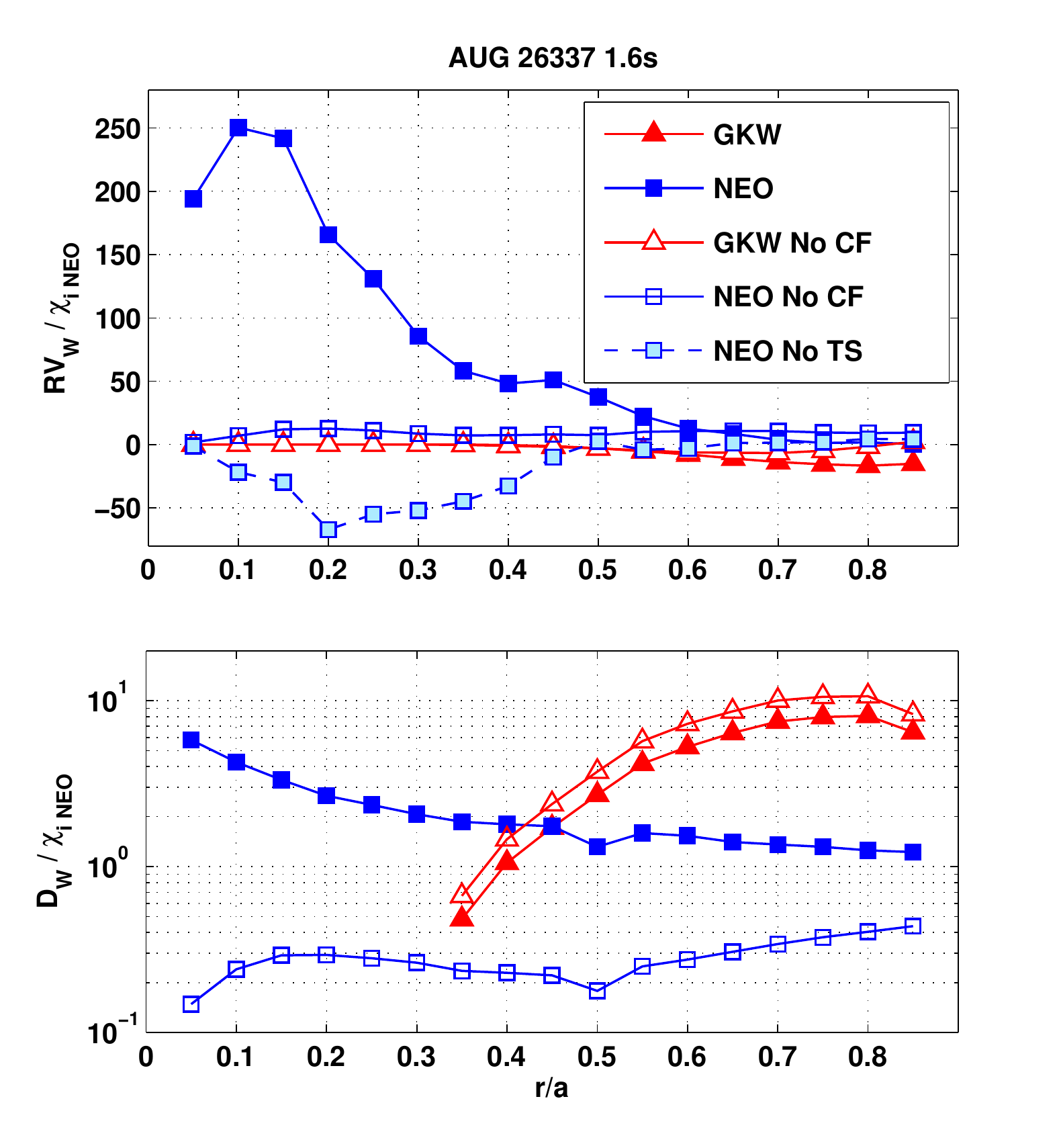}
\includegraphics[width=7truecm,trim=10 10 -3 8,clip=true]{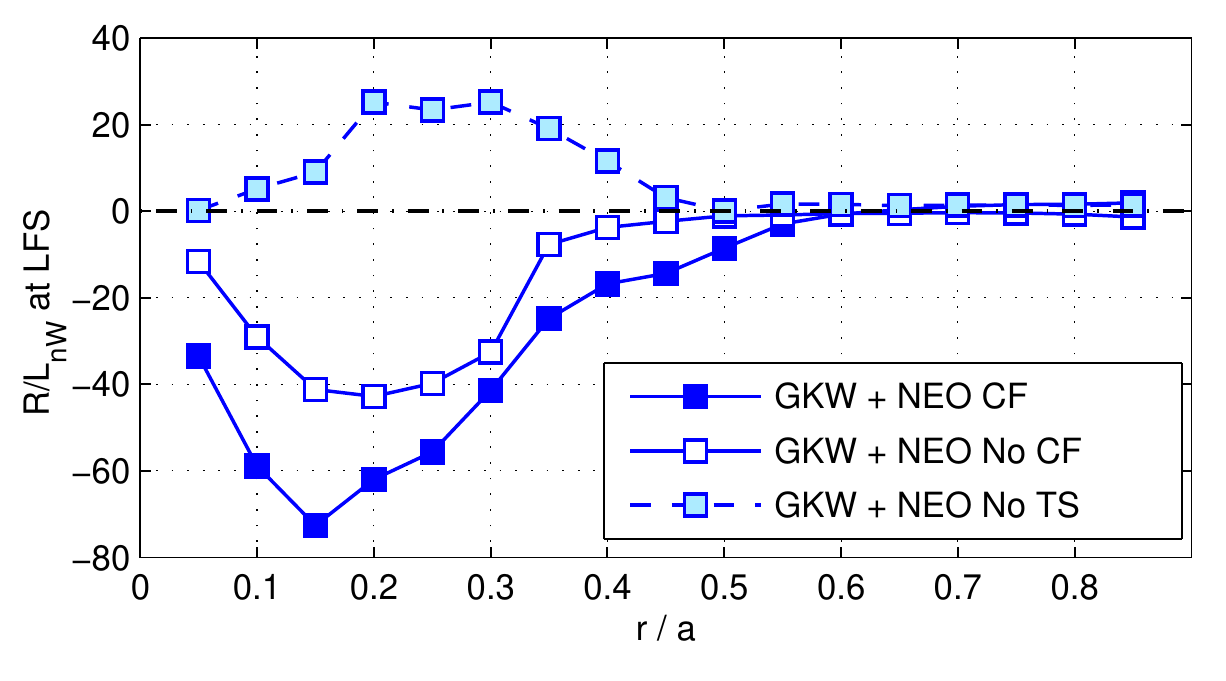}
\caption{Predicted W transport coefficients (LFS) and $R/L_{n_W}$ for AUG 26337 at $t = 1.6s$, with additional simulations
excluding the centrifugal force (No CF), and excluding neoclassical temperature screening (No TS).} 
\label{fig.aug_transport}
\end{center}
\end{figure}

The predicted transport coefficients in Fig. \ref{fig.aug_transport} show that these input lead to a strongly outward neoclassical convection over the whole profile, which dominates turbulent convection for $r/a < 0.7$.  For the diffusive transport, the turbulence dominates from $r/a > 0.45$.

\begin{figure}[tbh] 
\begin{center}
\includegraphics[width=4.3truecm,trim=70 110 550 20,clip=true]{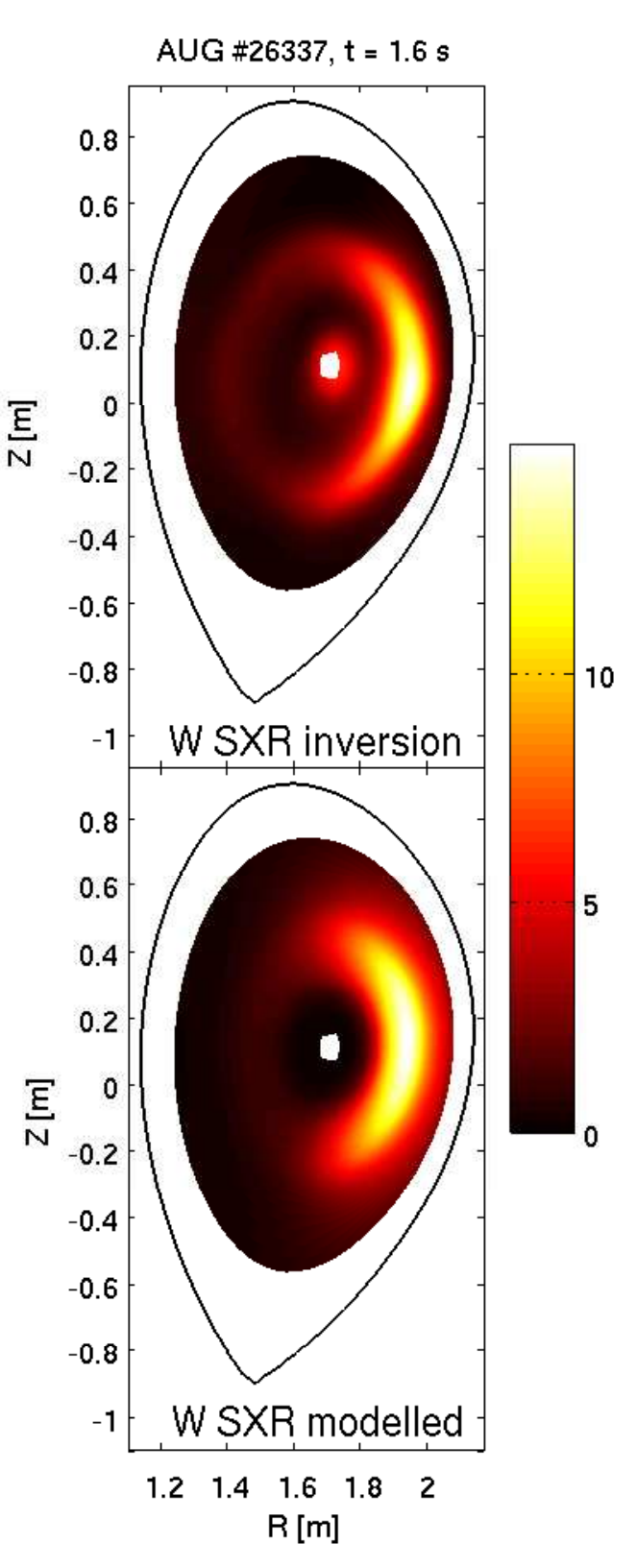}
\includegraphics[width=4.2truecm,trim=0 100 3 0,clip=true]{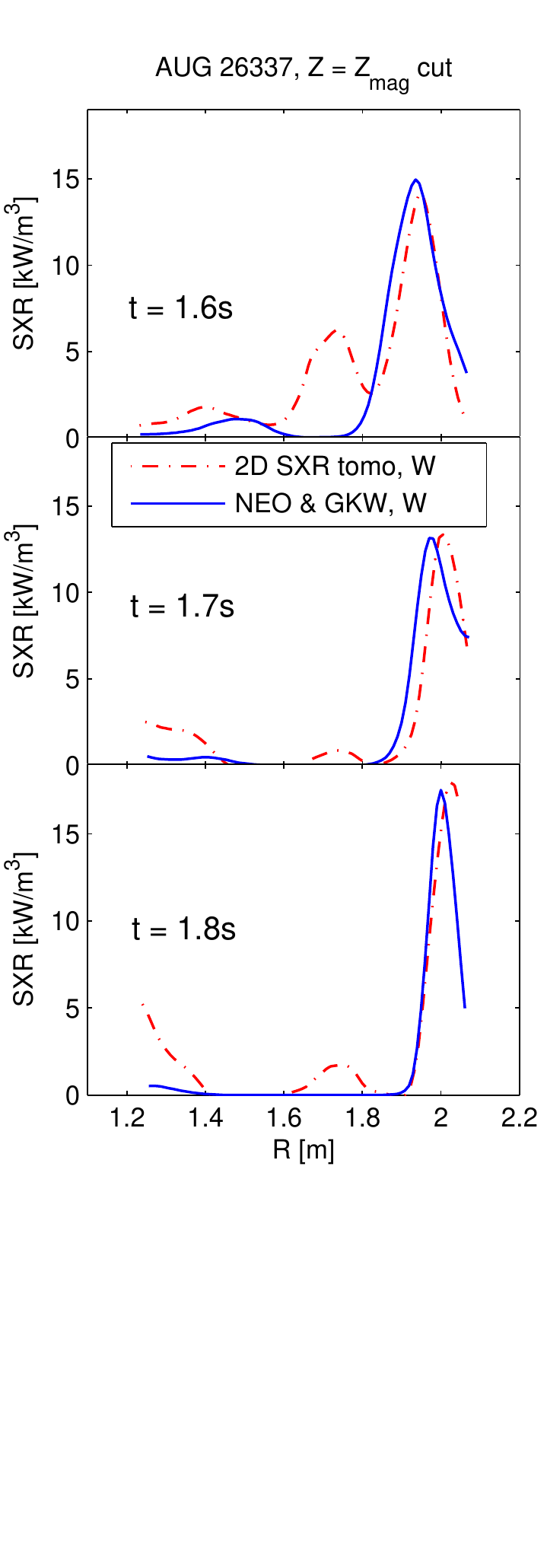}

\caption{Comparison of tomographic inversions of SXR emission with Brehmstrahlung subtracted (top left) 
and W SXR emission forward modelled from the predicted W distribution (bottom left) for t=1.6s of AUG 26337. The maximum value is used as a boundary condition in the modelled case to match the tomography. (right) The same data is cut horizontally through the magnetic axis (all three timeslices).
} 
\label{fig.aug_tomo}
\label{fig.aug_lfs}
\end{center}
\end{figure}

To validate these predictions, we compare predicted soft X-ray (SXR) emission (forward modelled from the predicted 2D W density) with SXR tomography with Bremsstrahlung radiation subtracted (for the modelled region only), under the assumption that W dominates the remaining emission.  Here, high quality SXR tomography is made possible by the high temperatures in this shot (in cooler AUG plasmas W emission falls below the filter cut-off at $\sim 2 keV$), and the recent application to AUG of the tomographic method described in Ref. \cite{odstrcil_modern_2012}.  In the outer half of the plasma, the comparison in Fig. \ref{fig.aug_tomo} shows agreement well within the uncertainties in both the radial and poloidal structure of the radiation, and provides an additional qualitative validation of the model.  Following the sensitivity discussion in Sec. \ref{sec.method}, uncertainties in the core $n_i$ profile 
are enough to account for the differences between prediction and tomography near the axis.  The disagreements at the HFS may be due to inaccuracies in the rotation measurement, giving overestimated predicted asymmetry, or contributions from lighter (less asymmetric) species still present in the tomography after the estimated Bremsstrahlung subtraction.  

To investigate the components of the model that are required, additional simulations are presented (see Fig. \ref{fig.aug_transport}):  When  CF effects are removed, the neoclassical transport drops by an order of magnitude and no longer dominates the turbulent transport, while the turbulent transport is relatively unaffected.  If instead the temperature screening is removed, (and CF effects are kept), the neoclassical transport remains enhanced but reverses sign, which would lead to strong central accumulation.  In removing either effect, the comparison to the tomography shows qualitative disagreement (not shown), indicating that both components are essential to the model.

To summarize, this case demonstrates that in advanced scenarios with strong rotation and 
strong temperature gradients but weak density gradients, 
neoclassical temperature screening alone
can be effective enough to trap W in the outer LFS region of the plasma, and turbulent transport
is not needed to avoid accumulation (indeed, for these heavy impurities, turbulent convection
will always struggle to compete with neoclassical convection in the core).

Thus, even in conditions of improved confinement, where neoclassical
transport dominates over the entire profile
there are conditions where turbulent transport is not needed to avoid core W accumulation.

\section{W transport under ICRH and NBI heating, JET baseline H-mode \label{sec.jet}}

In this section we model W in a pair of JET baseline H-modes in an ICRH power scan.
These shots are a follow-up to Ref. \cite{valisa_metal_2011}, where it was observed that central ICRH 
can reverse central impurity convection from inward to outward. 
The discharges have approximately the same total heating power; 14.7 MW NBI with 4.9 MW central ICRH in 85307, and 
19.1 MW NBI in 85308, and both include an H minority at $\sim 9\%$ concentration.  
In both cases the time selected for modelling was just prior to a sawtooth crash.

\begin{figure}[tbh] 
\begin{center}
\includegraphics[width=9.3truecm,trim=12 20 12 40,clip=true]{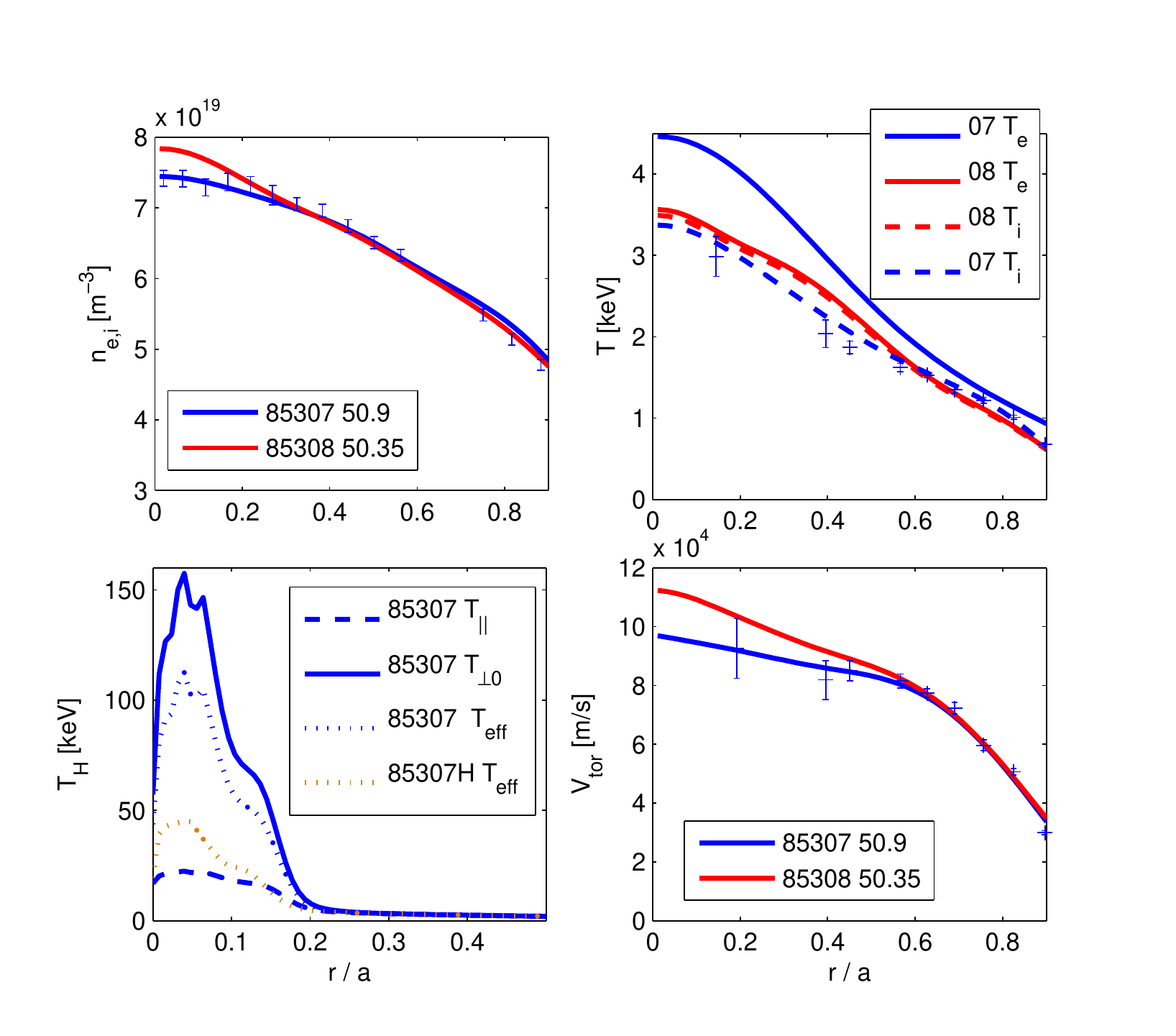}
\caption{Input profiles for the JET modelling, with indicitive selected data and error bars. (Bottom left) H minority temperatures produced by {\sc toric-ssfpql} for the ICRH case, inner radii only (the case 85307H is with half ICRH power).} 
\label{fig.jet_inputs}
\end{center}
\end{figure}

The model for poloidal asymmetry of W induced by anisotropic heating of the minority species (Sec. \ref{sec.theory})
requires inputs of $T_\parallel$ and $T_\perp$ for the minority species.  These are not measured directly,
but are simulated for 85307 using the the wave code {\sc toric} \cite{brambilla_numerical_1999} 
iteratively coupled \cite{bilato_simulations_2011} to the Fokker-Planck solver {\sc ssfpql} \cite{brambilla_quasi-linear_1994}.  The simulations were
performed for a pure plasma using the same kinetic profiles and full geometry as the {\sc gkw} + {\sc neo} simulations, with additional inputs of ICRH power, frequency and antenna phasing.  The minority temperature after the collisional slowing down is a nonlinear function of the absorbed power per particle.  Since central ion temperature measurements were not available, 
the ICRH power deposition profiles from {\sc toric-ssfpql} were also used as an input to an interpretive power balance in {\sc jetto}
to refine the central $T_i$ profile.  
These simulations do not include the interaction of NBI with ICRH, which may reduce the temperature and the anisotropy of the minority,
or finite orbit effects, which may widen the deposition profile and reduce the gradients. 

\begin{figure}[tbh] 
\begin{center}
\includegraphics[width=7truecm]{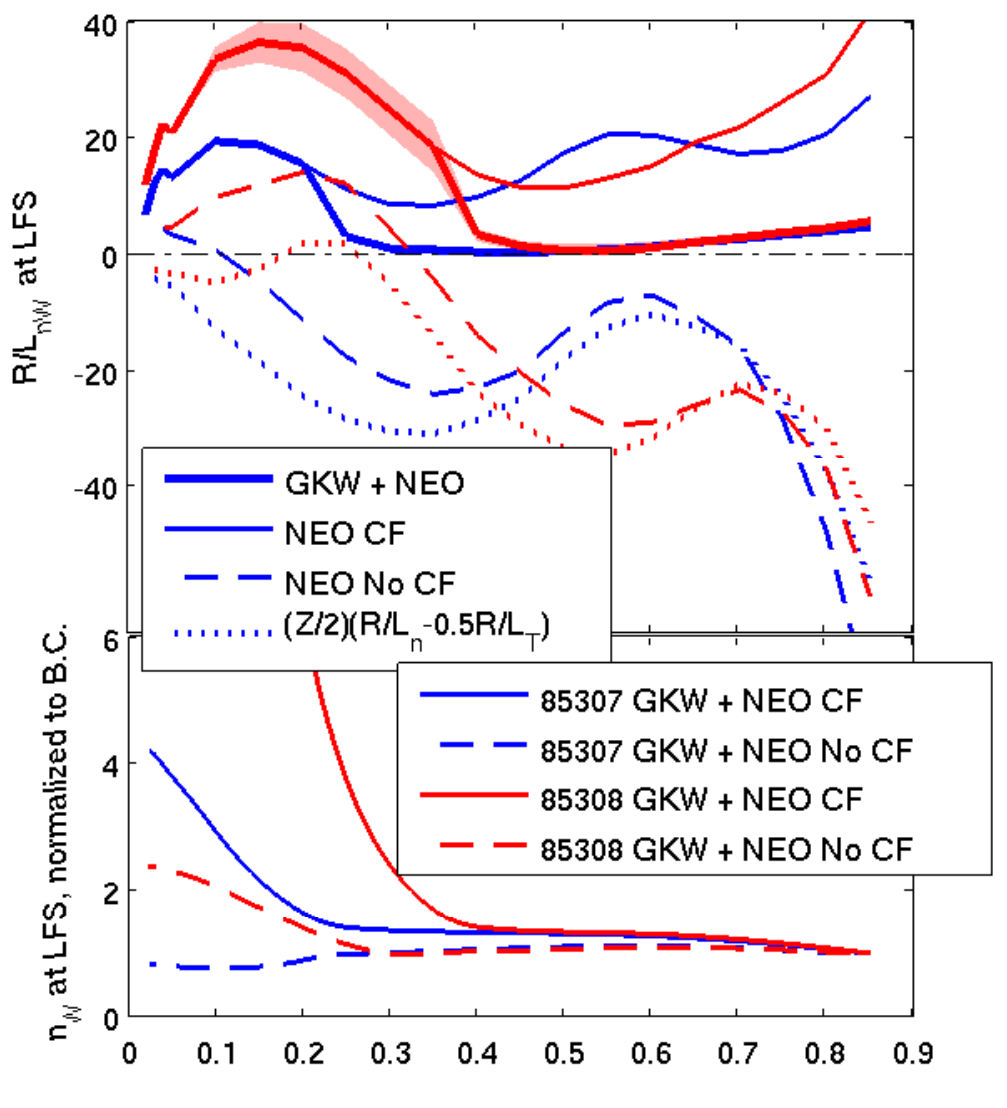}
\caption{Predicted $R/L_{n_W}$ (top) and integrated $n_W$ profiles (bottom) for JET 85308 w/o ICRH (red) and 85307 with ICRH (blue), with CF effects but no ICRH minority effects.  (top) For 85308, the red band indicates sensitivity to $\pm 10 \%$ changes in both $R/L_{n_i}$ and $R/L_{T_i}$ inputs. 
A simple analytic estimate of neoclassical peaking (dots) closely follows the {\sc neo} result w/o CF effects (dashes).} 
\label{fig.jet_cf_compare}
\end{center}
\end{figure}

\begin{figure*}[tbh] 
\begin{center}
\includegraphics[height=7.3truecm,trim=3 0 22 5,clip=true]{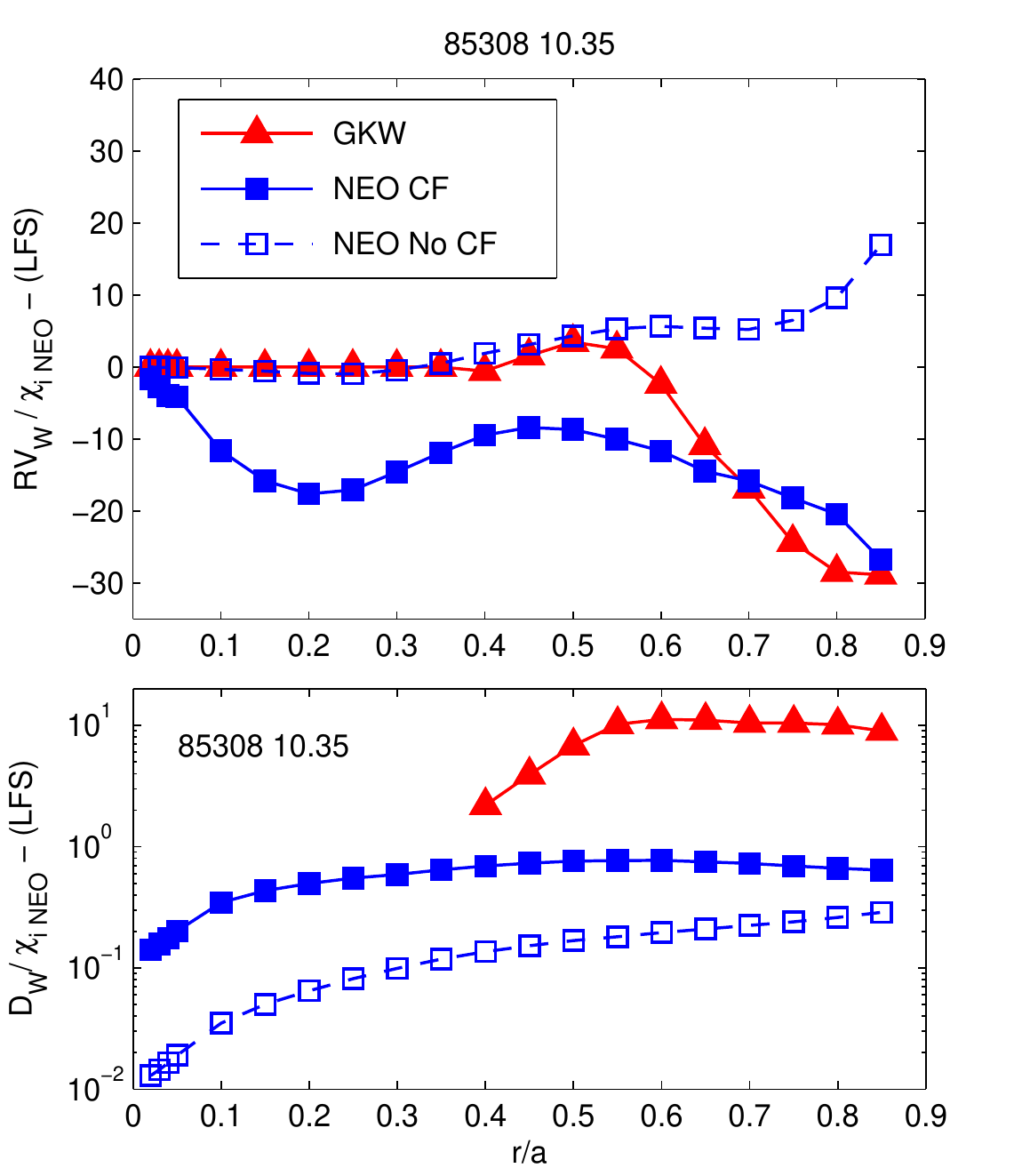}
\includegraphics[height=7.3truecm,trim=3 0 22 6,clip=true]{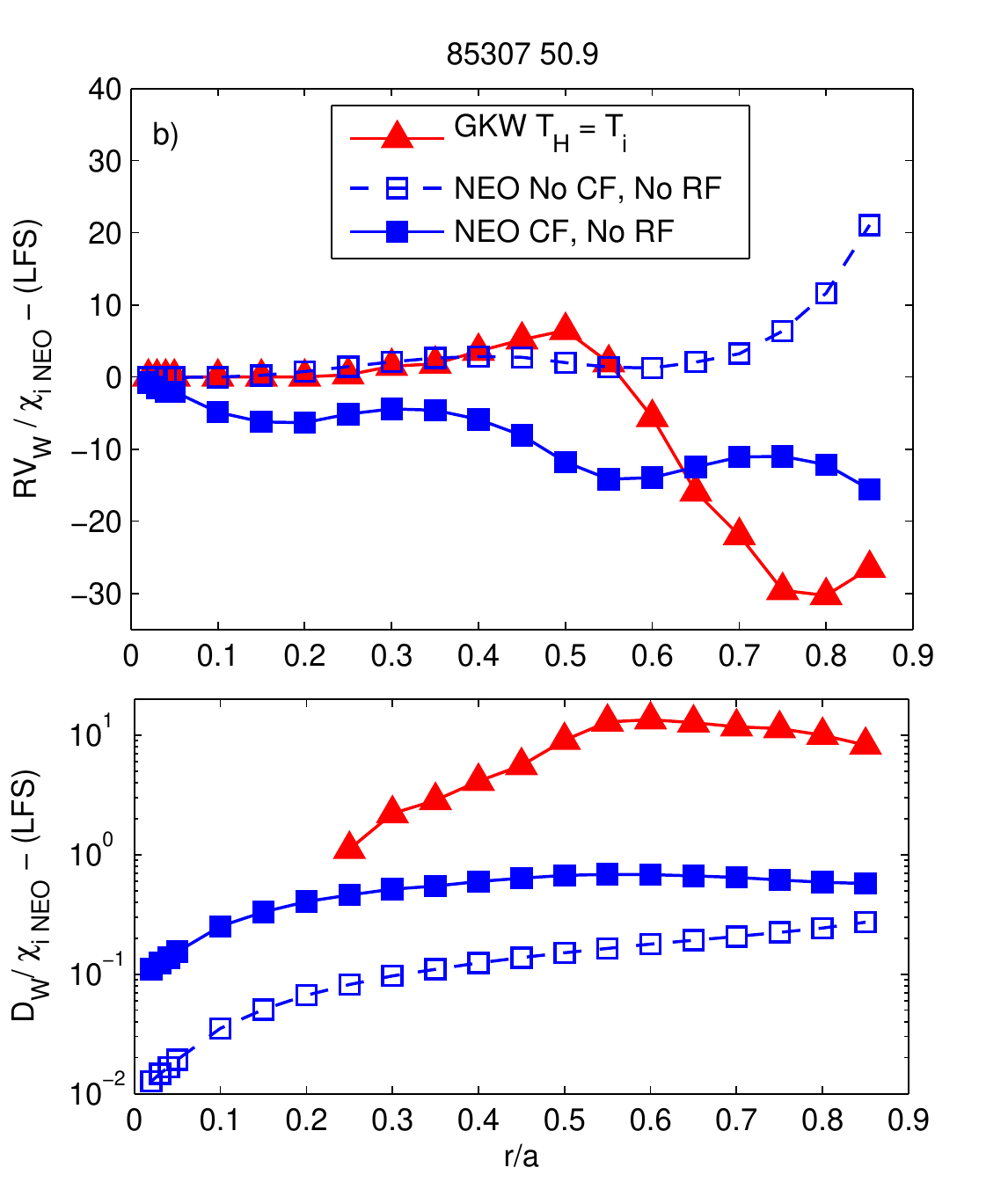}
\includegraphics[height=7.3truecm,trim=3 0 15 5,clip=true]{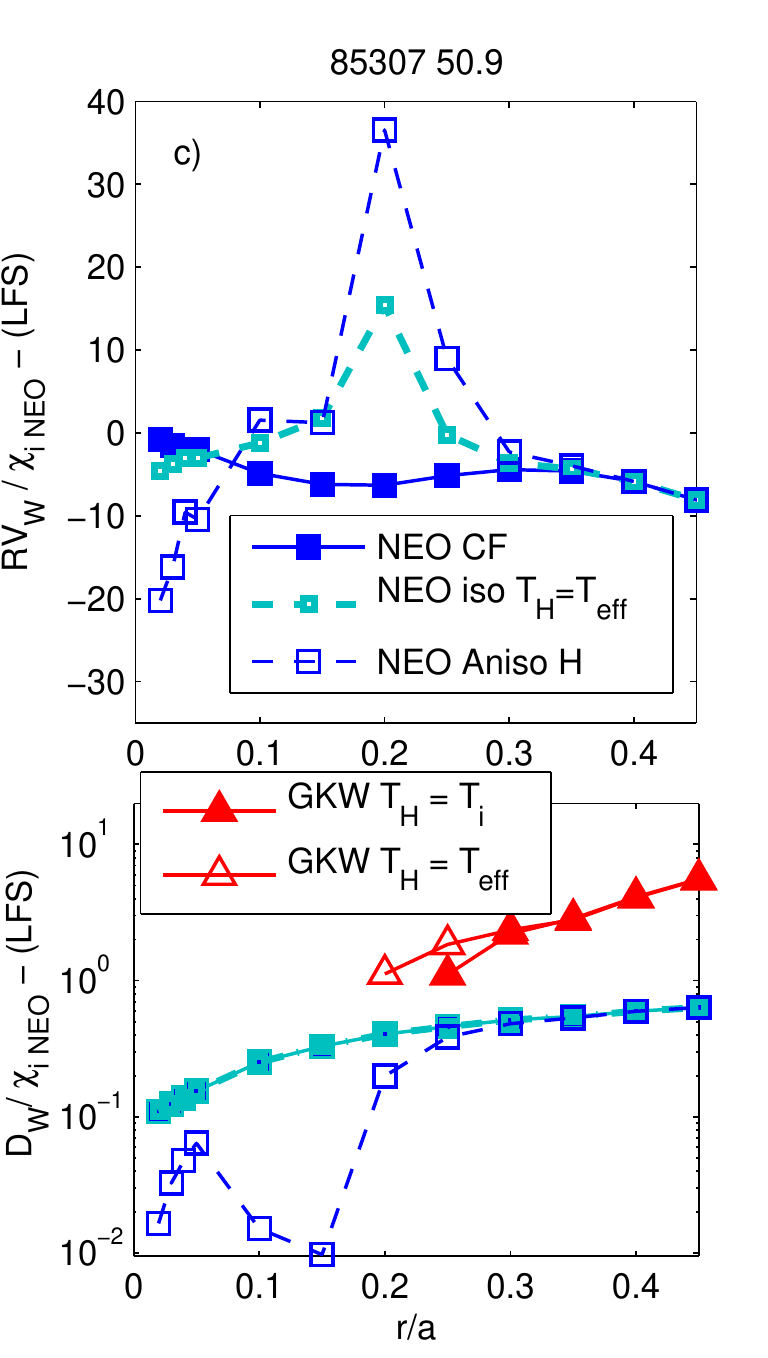}
\caption{Predicted W transport coefficients for the JET cases.  
For 85307, results include CF effects only ($T_H = T_{i}$, middle), CF + heated isotropic minority ($T_H = T_{\rm eff}$, right), and CF + heated anisotropic minority (Aniso H, {\sc neo} only, right).}
\label{fig.jet_dv}
\end{center}
\end{figure*}

The modelling inputs are shown in Fig. \ref{fig.jet_inputs}.  Discharge 85307 has hotter electrons in the core, since more ICRH power goes to the electrons, but $T_i$, which determines the W transport, is similar.  The higher rotation and more peaked density in 85308 are the key differences which determine the different predictions in Figs. \ref{fig.jet_cf_compare} and \ref{fig.jet_dv}a,b.  Also shown in Fig. \ref{fig.jet_inputs} are the anisotropic H minority temperatures produced from {\sc toric-ssfpql}.

In the first stage of modelling, the simulations included CF effects only (with $T_H= T_D$), as in the previous section.
Both predicted profiles show central W peaking (Fig. \ref{fig.jet_cf_compare}), 
enhanced by CF effects due to the reduction in temperature screening 
relative to the pinch.  The CF effects have a slightly larger impact in 85308 due to the larger rotation (Fig. \ref{fig.jet_dv}a,b). 
Without CF effects, the {\sc neo}-only $R/L_{n_W}$ closely follows a simple neoclassical estimate $\propto R/L_{n_i} - 0.5 R/L_{T_i}$ 
for the PS regime; already here we see that 85308, without ICRH, shows stronger central peaking simply due to its
more peaked density profile and stronger rotation.  (The reasons for the more peaked density profile in 85308 are not investigated in this work, but are
likely due to less central turbulence offsetting the Ware pinch, and an increased particle source from NBI \cite{angioni_particle_2009,giroud_this_2014}.)

\begin{figure}[tbh] 
\begin{center}
\includegraphics[height=5truecm,trim=5 0 12 0,clip=true]{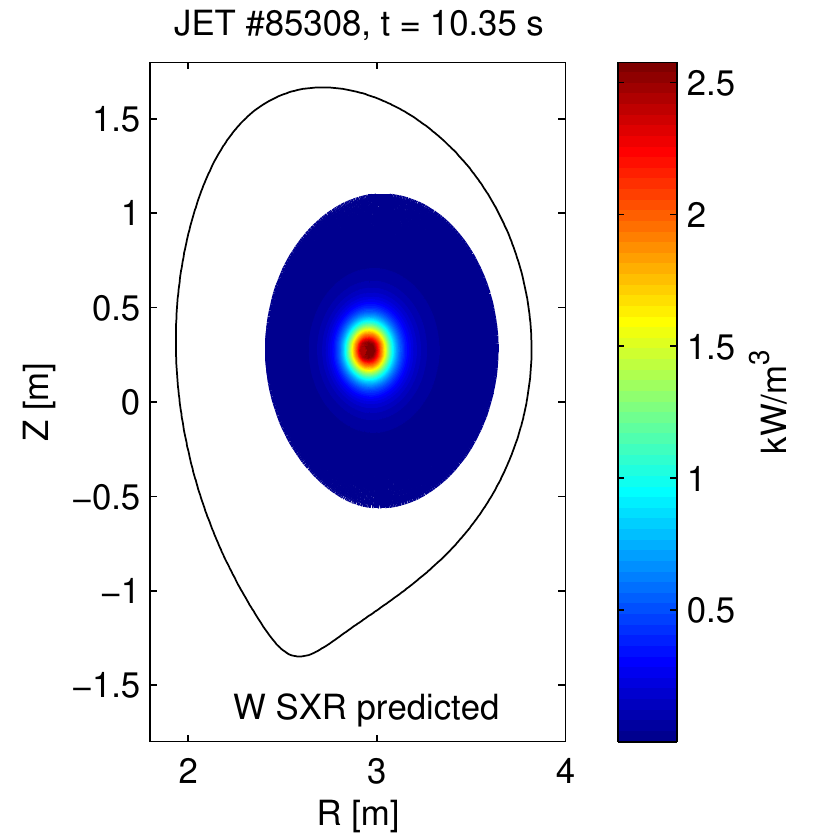}
\includegraphics[height=5truecm,trim=42 0 12 0,clip=true]{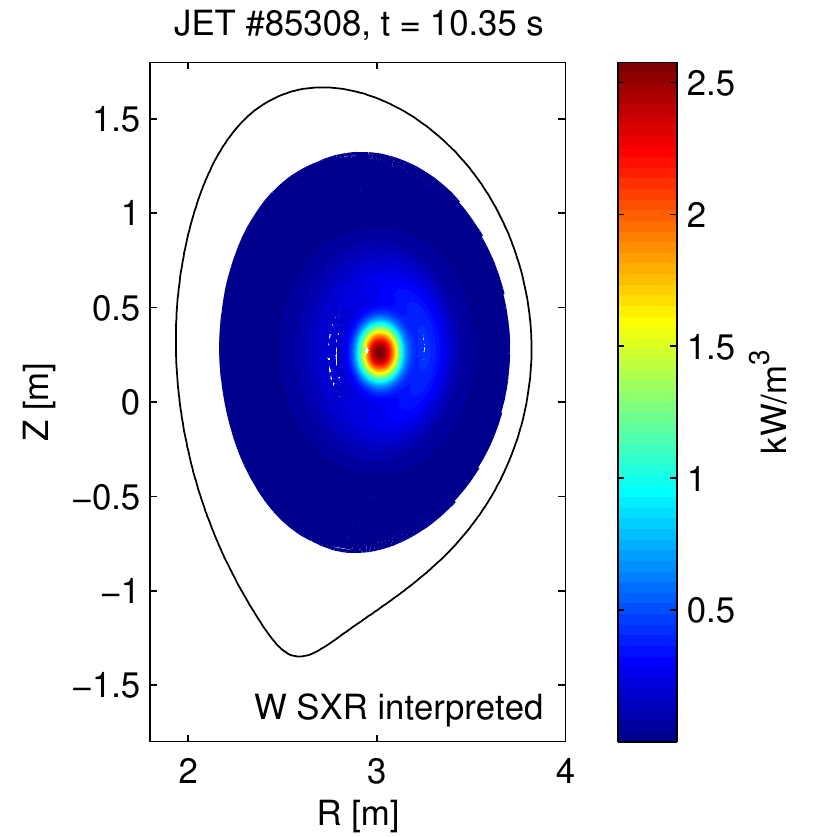}
\caption{Comparison of predicted and interpreted SXR emission from W for JET 85308 (NBI only).
The predicted scale is matched to the interpreted value at the central maximum.} 
\label{fig.jet_tomo08}
\end{center}
\end{figure}

For 85308, without ICRH, the 2D W SXR prediction shows good qualitative agreement with the interpreted 
SXR tomography (Fig \ref{fig.jet_tomo08}) using the tool developed for Ref. \cite{putterich_tungsten_2012,putterich_observations_2013}.
For the reasons discussed in Sec III (particularly the presence of large sawteeth), 
the comparison does not show the same level of quantitative agreement over the full profile 
as the AUG results above, but nevertheless demonstrates that, 
for the case without ICRH, the model including CF effects correctly predicts W accumulation.

\begin{figure*}[tbh] 
\begin{center}
\includegraphics[height=5.3truecm,trim=5 0 5 0,clip=true]{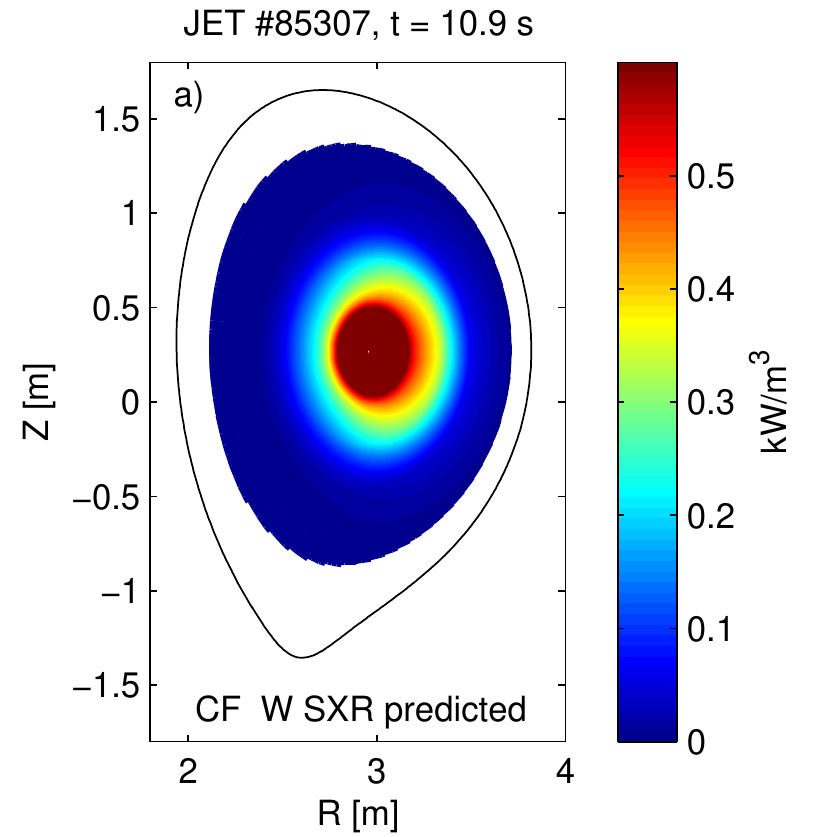}
\includegraphics[height=5.3truecm,trim=42 0 12 0,clip=true]{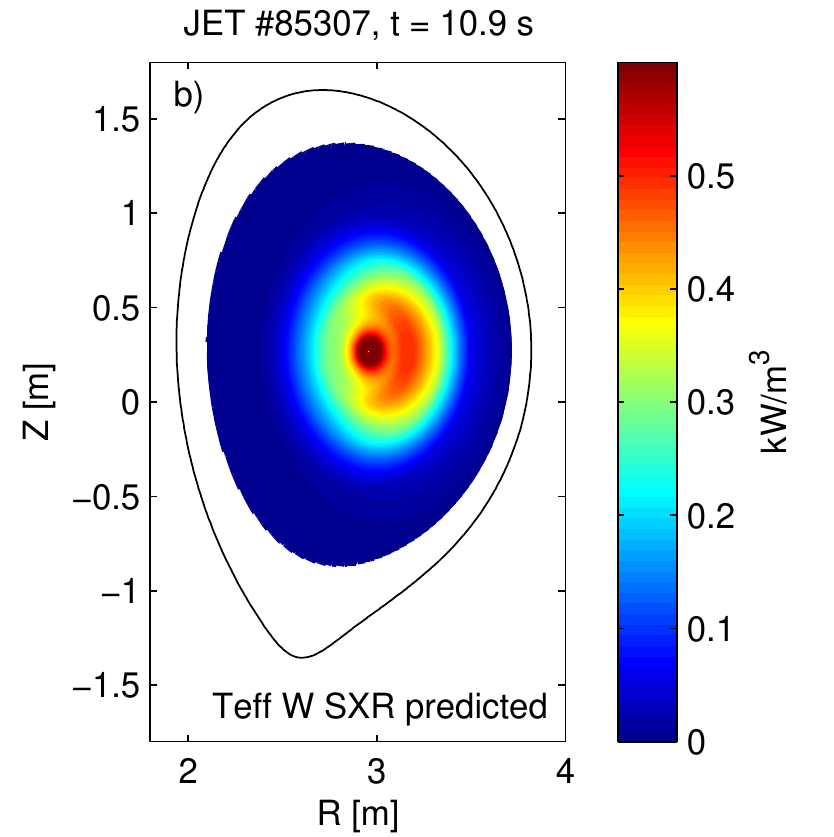}
\includegraphics[height=5.3truecm,trim=42 0 12 0,clip=true]{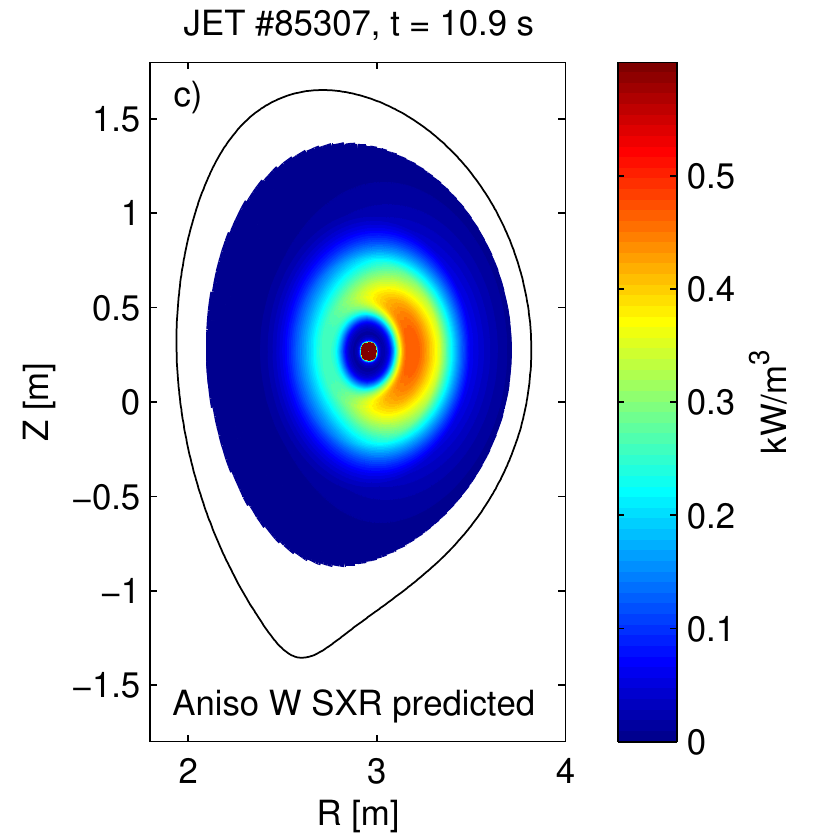}
\includegraphics[height=5.3truecm,trim=42 0 12 0,clip=true]{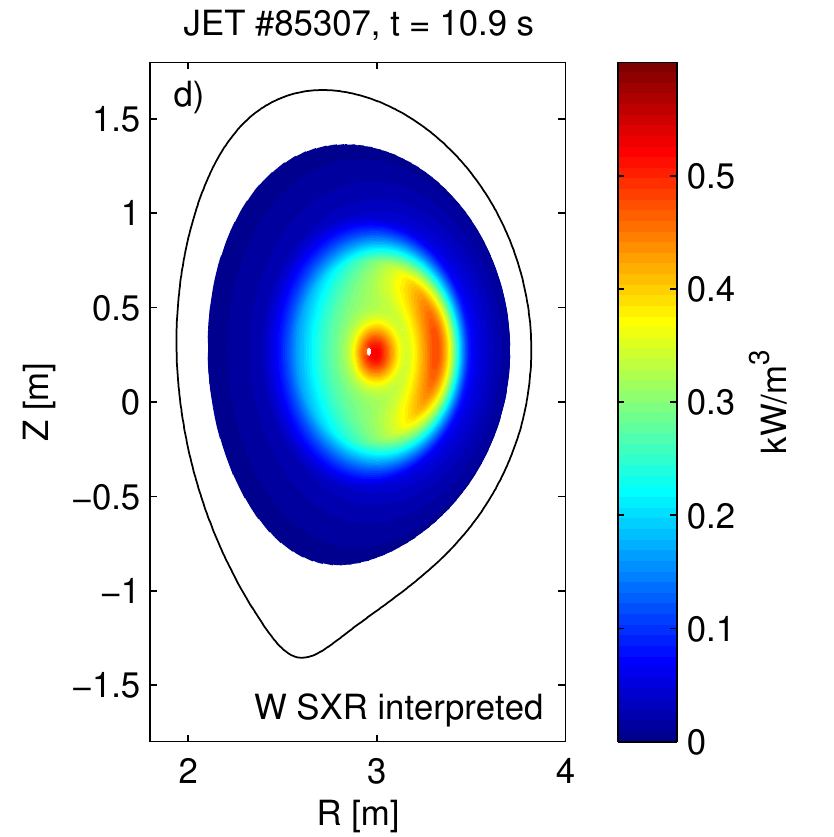}
\caption{Comparison of predicted and interpreted (d) SXR emission from W for JET 85307 (NBI + ICRH).  From left to right
the predictions include: CF effects only (a), CF effects with heated isotropic minority (b), and CF effects with heated anisotropic minority ({\sc neo} only) (c). 
The predictions are 
matched to the interpreted value at $r/a=0.35$, in the outer LFS maximum.} 
\label{fig.jet_tomo07}
\end{center}
\end{figure*}

In contrast, for 85307, with CF effects only, the centrally peaked density profile   
does not agree with the tomography (Fig \ref{fig.jet_tomo07} a vs d), 
and indicates a possible missing piece in the modelling, motivating 
the progressive inclusion of the minority heating effects (Fig. \ref{fig.jet_dv}c): 

First, the effective isotropic minority temperature from {\sc toric-ssfpql} is added 
to the minority species 
which is kept isotropic with
$T_{\rm eff} = (T_\parallel + 2 T_{\perp R0} ) / 3$.  For the {\sc gkw} simulations, the increased minority temperature gradient shifts the stability boundary slightly inward, but the impact is much larger on the neoclassical transport. The heated minority 
does not change the neoclassical diffusivity (Fig. \ref{fig.jet_dv}), but switches the neoclassical convection to strongly
outward in the region of the ICRH absorption ($0.1 < r/a < 0.3$), due to an additional temperature screening from collisions between W and H.  
Notably, this additional temperature screening becomes negative at $r/a < 0.1$, in exactly the region where $R/L_{T_{\rm eff}} < 0$ for the minority.
The ion-impurity friction which drives temperature screening \cite{angioni_neoclassical_2014} scales as $\propto n_i T_i \nu_{iZ} R / L_{T_{i}}$.  For the H-W and D-W collisions with $Z_W = 46$, these parameters are given in Table I, and demonstrate that the minority H 
contributes a screening of the same order of magnitude as the bulk D at $r/a = 0.2 - 0.25$, effectively doubling the strength of the screening.  We note that at the very high $T_H$,
 the minority collisions decouple (in both Table I and Fig. \ref{fig.jet_dv}), and 
the maximum minority screening effect is not at the ICRH resonance at $r/a = 0.07$, but at the edges of the heated region.  For this reason, this additional screening is very sensitive to the exact details of the minority temperature profile from {\sc toric-ssfpql}.

\begin{table}[t!]
\squeezetable
\begin{footnotesize}
\begin{tabular}{|l|l|l|l|l|l|l|}
\hline
Ion & $r/a$ & $n_i [10^{19} m^{-3}]$ & $T_i [keV]$ & $\frac{R}{L_{T_{i}}}$ & $\frac{\nu_{iW}}{{v_{\rm th,i} / R}}$ & $n_i T_i \frac{\nu_{iW}}{{v_{\rm th,i} / R}} \frac{R}{L_{T_{i}}}$ \\ 
\hline
H & 0.10  &   0.664 &  63.5  &  30.7 &   0.0016 &  2.1  \\
H & 0.15  &   0.658 &  45.2  &  50.3 &   0.0032 &  4.7  \\
H & 0.20  &   0.650 &  7.76  &  97.8 &   0.1038 &  51.2  \\
H & 0.25  &   0.642 &  3.48  &  37.5 &   0.5156 &  43.3  \\
\hline
D & 0.10  &   6.72 &   3.26  &  2.16  &  0.61 &  29.0 \\
D & 0.15  &   6.65 &   3.13  &  2.99  &  0.66 &  41.6 \\
D & 0.20  &   6.57 &   2.97  &  3.76  &  0.70 &  52.1 \\
D & 0.25  &   6.49 &   2.79  &  4.21  &  0.80 &  61.3 \\
\hline
\end{tabular}
\end{footnotesize}
\caption{Comparision of parameters in ion-W screening for collisions with H and D ions.  For readable numbers, $n_W = 10^{19} m^{-3}$ (arbitrary) was used for $\nu_{iW}$.} 
\end{table}

\begin{figure}[tbh] 
\begin{center}
\includegraphics[width=6truecm,trim=0 20 0 5,clip=true]{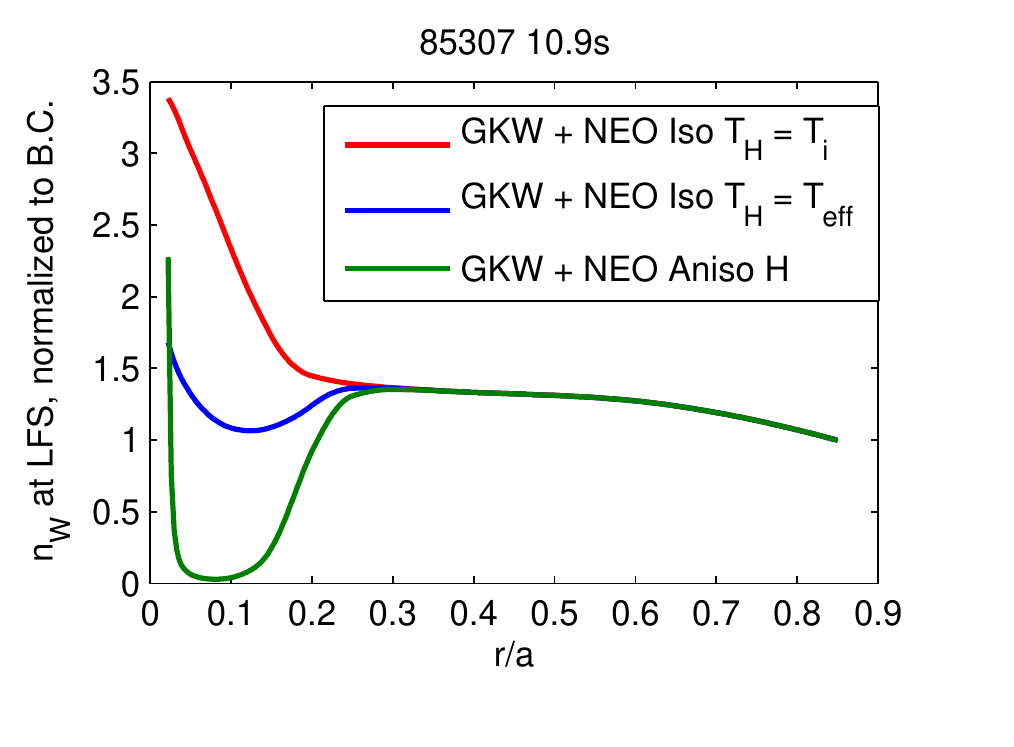}
\caption{Comparison of LFS predicted profiles for JET 85307 with ICRH minority effects (labels as in Fig. \ref{fig.jet_dv}).} 
\label{fig.jet_rf_compare}
\end{center}
\end{figure}
        
Second, the minority is made anisotropic using the simulated $T_\parallel, T_\perp$ as inputs to the model of Eq. \ref{eq.bilato}. 
The result (Fig. \ref{fig.jet_dv}c) is a strong reduction in neoclassical diffusivity, as expected from Sec. II, due 
to the reduction of the $P_A$ factor (Fig. \ref{fig.pol_factors}).
Additionally, the minority temperature screening effect is strongly enhanced in the regions where 
$P_A \gg 1$.  In these regions, the CF asymmetry dominates, producing LFS W localisation,
so both W and H are localised on the LFS, increasing their local collision frequency, 
and amplifying the minority temperature screening effect 
(the details of this synergy remain to be clarified).

The end result of the additional temperature screening is to significantly
flatten the central W profile (Fig. \ref{fig.jet_rf_compare}) 
with the reversal of the minority temperature screening even causing a second, central, peak 
in qualitative agreement with the tomography (Fig. \ref{fig.jet_tomo07}b,d).
The effects of the anisotropy (Fig. \ref{fig.jet_tomo07}c) appear to overly exaggerate the dip in $n_W$ close to the axis.
Given the additional effects not included {\sc toric} described in Sec. \ref{sec.method}, both minority effects
in our results should be considered an upper estimate.  In sensitivity tests with half ICRH power
we observe that the minority effects are qualitatively robust, but change quantitatively depending
on the inputs from {\sc toric-ssfpql}.

We note that the ICRH minority effects described here are consistent with the reversal of the convection described
in \cite{valisa_metal_2011}; future work will compare $D_{\rm Mo}$ and $V_{\rm Mo}$ predictions to laser blow off fits, and should include these transport coefficients in time evolution of W integrated modelling.   The minority screening effect combined with the anisotropy may also explain the strong Mo peaking at $r/a = 0.55$ in Ref. \cite{mollen_impurity_2014}; in that case, if $P_A$ is negative due to the HFS impurity localisation, all neoclassical transport including the minority screening would reverse; we leave confirmation for future work.  These effects should also be quantified for NBI fast ions.

\section{Conclusions}

In this work, we have modelled turbulent and neoclassical heavy impurity transport  
using theory-based numerical tools ({\sc gkw} and {\sc neo} respectively) with comprehensive treatment of poloidal asymmetries, 
to predict core W distributions in JET and AUG. 

In the ASDEX-Upgrade improved H-mode with current overshoot, the flat density profiles mean that
neoclassical temperature screening is sufficient to prevent accumulation and trap W in the outer half of the plasma. 
Centrifugal effects enhance neoclassical transport by an order of magnitude such that it dominates impurity 
turbulent transport over most of the plasma radius.

In JET H-modes with ICRH, strong minority heating enhances neoclassical impurity temperature screening, 
and reverses the convection in the region of the ICRH (in agreement with Ref. \cite{valisa_metal_2011}).
In addition, the anisotropy-induced poloidal asymmetry reduces neoclassical impurity diffusivity, 
and the minority-impurity temperature screening may be enhanced when both species are localised at the LFS.
These effects are complementary to flatter density profiles in ICRH plasmas, and help to prevent central W accumulation.

Comparing our predictions with tomographic inversions from soft X-ray measurements,
we have demonstrated further validation of these models over a greater range of plasma conditions. 
This validation re-emphasizes that poloidal asymmetries are an essential ingredient for accurate modelling of (particularly neoclassical) heavy impurity transport. Additionally, we have shown that the temperature gradients of externally heated species can contribute significantly to impurity temperature screening, and should also be included in neoclassical modelling.  Experiments with off-axis heating may be able to further probe and isolate these effects.

\vspace{5pt}

\begin{acknowledgments}
FJC would like to thank Colin Roach for helpful comments, and Arthur Peeters for many helpful discussions and for making the {\sc gkw} code available.  
This project has received funding from the European Union's Horizon 2020 research and innovation programme
under grant agreement number 633053, and by the RCUK Energy Programme [grant number EP/I501045], and the Max Planck Institute.  The views and opinions expressed herein do not necessarily reflect those of the European Commission.

\end{acknowledgments}

\bibliographystyle{my_doi2} 
\bibliography{eps_2014}

\begin{thebibliography}{49}
\expandafter\ifx\csname natexlab\endcsname\relax\def\natexlab#1{#1}\fi
\expandafter\ifx\csname bibnamefont\endcsname\relax
  \def\bibnamefont#1{#1}\fi
\expandafter\ifx\csname bibfnamefont\endcsname\relax
  \def\bibfnamefont#1{#1}\fi
\expandafter\ifx\csname citenamefont\endcsname\relax
  \def\citenamefont#1{#1}\fi
\expandafter\ifx\csname url\endcsname\relax
  \def\url#1{\texttt{#1}}\fi
\expandafter\ifx\csname urlprefix\endcsname\relax\def\urlprefix{URL }\fi
\providecommand{\bibinfo}[2]{#2}
\providecommand{\eprint}[2][]{\url{#2}}

\bibitem[{\citenamefont{Shimada et~al.}(2009)\citenamefont{Shimada, Pitts,
  Loarte et~al.}}]{shimada_iter_2009}
\bibinfo{author}{\bibfnamefont{M.}~\bibnamefont{Shimada}},
  \bibinfo{author}{\bibfnamefont{R.}~\bibnamefont{Pitts}},
  \bibinfo{author}{\bibfnamefont{A.}~\bibnamefont{Loarte}},
  \bibnamefont{et~al.}, \bibinfo{journal}{J. Nucl. Mat.}
  \href{http://dx.doi.org/10.1016/j.jnucmat.2009.01.113}{\textbf{\bibinfo{volu%
me}{390{\textendash}391}}, \bibinfo{pages}{282}} (\bibinfo{year}{2009}).

\bibitem[{\citenamefont{Hinton and Wong}(1985)}]{hinton_neoclassical_1985}
\bibinfo{author}{\bibfnamefont{F.~L.} \bibnamefont{Hinton}} \bibnamefont{and}
  \bibinfo{author}{\bibfnamefont{S.~K.} \bibnamefont{Wong}},
  \bibinfo{journal}{Phys. Fluids}
  \href{http://dx.doi.org/10.1063/1.865350}{\textbf{\bibinfo{volume}{28}},
  \bibinfo{pages}{3082}} (\bibinfo{year}{1985}).

\bibitem[{\citenamefont{Wesson}(1997)}]{wesson_poloidal_1997}
\bibinfo{author}{\bibfnamefont{J.}~\bibnamefont{Wesson}},
  \bibinfo{journal}{Nucl. Fusion}
  \href{http://dx.doi.org/10.1088/0029-5515/37/5/I01}{\textbf{\bibinfo{volume}%
{37}}, \bibinfo{pages}{577}} (\bibinfo{year}{1997}).

\bibitem[{\citenamefont{Chang}(1983)}]{chang_enhancement_1983}
\bibinfo{author}{\bibfnamefont{C.~S.} \bibnamefont{Chang}},
  \bibinfo{journal}{Phys. Fluids}
  \href{http://dx.doi.org/10.1063/1.864396}{\textbf{\bibinfo{volume}{26}},
  \bibinfo{pages}{2140}} (\bibinfo{year}{1983}).

\bibitem[{\citenamefont{Helander}(1998{\natexlab{a}})}]{helander_neoclassical_%
1998}
\bibinfo{author}{\bibfnamefont{P.}~\bibnamefont{Helander}},
  \bibinfo{journal}{Phys. Plasmas}
  \href{http://dx.doi.org/10.1063/1.872629}{\textbf{\bibinfo{volume}{5}},
  \bibinfo{pages}{1209}} (\bibinfo{year}{1998}{\natexlab{a}}).

\bibitem[{\citenamefont{Helander}(1998{\natexlab{b}})}]{helander_bifurcated_19%
98}
\bibinfo{author}{\bibfnamefont{P.}~\bibnamefont{Helander}},
  \bibinfo{journal}{Phys. Plasmas}
  \href{http://dx.doi.org/10.1063/1.873121}{\textbf{\bibinfo{volume}{5}},
  \bibinfo{pages}{3999}} (\bibinfo{year}{1998}{\natexlab{b}}).

\bibitem[{\citenamefont{Romanelli and
  Ottaviani}(1998)}]{romanelli_effects_1998}
\bibinfo{author}{\bibfnamefont{M.}~\bibnamefont{Romanelli}} \bibnamefont{and}
  \bibinfo{author}{\bibfnamefont{M.}~\bibnamefont{Ottaviani}},
  \bibinfo{journal}{Plasma Phys. Control. Fusion}
  \href{http://dx.doi.org/10.1088/0741-3335/40/10/007}{\textbf{\bibinfo{volume%
}{40}}, \bibinfo{pages}{1767}} (\bibinfo{year}{1998}).

\bibitem[{\citenamefont{F{\"u}l{\"o}p and
  Helander}(1999)}]{fulop_nonlinear_1999}
\bibinfo{author}{\bibfnamefont{T.}~\bibnamefont{F{\"u}l{\"o}p}}
  \bibnamefont{and} \bibinfo{author}{\bibfnamefont{P.}~\bibnamefont{Helander}},
  \bibinfo{journal}{Phys. Plasmas}
  \href{http://dx.doi.org/doi:10.1063/1.873593}{\textbf{\bibinfo{volume}{6}},
  \bibinfo{pages}{3066}} (\bibinfo{year}{1999}).

\bibitem[{\citenamefont{F{\"u}l{\"o}p et~al.}(2002)\citenamefont{F{\"u}l{\"o}p,
  Helander, and Catto}}]{fulop_effect_2002}
\bibinfo{author}{\bibfnamefont{T.}~\bibnamefont{F{\"u}l{\"o}p}},
  \bibinfo{author}{\bibfnamefont{P.}~\bibnamefont{Helander}}, \bibnamefont{and}
  \bibinfo{author}{\bibfnamefont{P.~J.} \bibnamefont{Catto}},
  \bibinfo{journal}{Phys. Rev. Lett.}
  \href{http://dx.doi.org/10.1103/PhysRevLett.89.225003}{\textbf{\bibinfo{volu%
me}{89}}, \bibinfo{pages}{225003}} (\bibinfo{year}{2002}).

\bibitem[{\citenamefont{Parisot et~al.}(2008)\citenamefont{Parisot, Guirlet,
  Bourdelle et~al.}}]{parisot_experimental_2008}
\bibinfo{author}{\bibfnamefont{T.}~\bibnamefont{Parisot}},
  \bibinfo{author}{\bibfnamefont{R.}~\bibnamefont{Guirlet}},
  \bibinfo{author}{\bibfnamefont{C.}~\bibnamefont{Bourdelle}},
  \bibnamefont{et~al.}, \bibinfo{journal}{Plasma Phys. Control. Fusion}
  \href{http://dx.doi.org/10.1088/0741-3335/50/5/055010}{\textbf{\bibinfo{volu%
me}{50}}, \bibinfo{pages}{055010}} (\bibinfo{year}{2008}).

\bibitem[{\citenamefont{Valisa et~al.}(2011)\citenamefont{Valisa, Carraro,
  Predebon et~al.}}]{valisa_metal_2011}
\bibinfo{author}{\bibfnamefont{M.}~\bibnamefont{Valisa}},
  \bibinfo{author}{\bibfnamefont{L.}~\bibnamefont{Carraro}},
  \bibinfo{author}{\bibfnamefont{I.}~\bibnamefont{Predebon}},
  \bibnamefont{et~al.}, \bibinfo{journal}{Nucl. Fusion}
  \href{http://dx.doi.org/10.1088/0029-5515/51/3/033002}{\textbf{\bibinfo{volu%
me}{51}}, \bibinfo{pages}{033002}} (\bibinfo{year}{2011}).

\bibitem[{\citenamefont{Ingesson et~al.}(2000)\citenamefont{Ingesson, Chen,
  Helander et~al.}}]{ingesson_comparison_2000}
\bibinfo{author}{\bibfnamefont{L.~C.} \bibnamefont{Ingesson}},
  \bibinfo{author}{\bibfnamefont{H.}~\bibnamefont{Chen}},
  \bibinfo{author}{\bibfnamefont{P.}~\bibnamefont{Helander}},
  \bibnamefont{et~al.}, \bibinfo{journal}{Plasma Phys. Control. Fusion}
  \href{http://dx.doi.org/10.1088/0741-3335/42/2/308}{\textbf{\bibinfo{volume}%
{42}}, \bibinfo{pages}{161}} (\bibinfo{year}{2000}).

\bibitem[{\citenamefont{Reinke et~al.}(2012)\citenamefont{Reinke, Hutchinson,
  Rice et~al.}}]{reinke_poloidal_2012}
\bibinfo{author}{\bibfnamefont{M.~L.} \bibnamefont{Reinke}},
  \bibinfo{author}{\bibfnamefont{I.~H.} \bibnamefont{Hutchinson}},
  \bibinfo{author}{\bibfnamefont{J.~E.} \bibnamefont{Rice}},
  \bibnamefont{et~al.}, \bibinfo{journal}{Plasma Phys. Control. Fusion}
  \href{http://dx.doi.org/10.1088/0741-3335/54/4/045004}{\textbf{\bibinfo{volu%
me}{54}}, \bibinfo{pages}{045004}} (\bibinfo{year}{2012}).

\bibitem[{\citenamefont{Bilato et~al.}(2014)\citenamefont{Bilato, Maj, and
  Angioni}}]{bilato_modelling_2014}
\bibinfo{author}{\bibfnamefont{R.}~\bibnamefont{Bilato}},
  \bibinfo{author}{\bibfnamefont{O.}~\bibnamefont{Maj}}, \bibnamefont{and}
  \bibinfo{author}{\bibfnamefont{C.}~\bibnamefont{Angioni}},
  \bibinfo{journal}{Nucl. Fusion}
  \href{http://dx.doi.org/10.1088/0029-5515/54/7/072003}{\textbf{\bibinfo{volu%
me}{54}}, \bibinfo{pages}{072003}} (\bibinfo{year}{2014}).

\bibitem[{\citenamefont{F{\"u}l{\"o}p and Moradi}(2011)}]{fulop_effect_2011}
\bibinfo{author}{\bibfnamefont{T.}~\bibnamefont{F{\"u}l{\"o}p}}
  \bibnamefont{and} \bibinfo{author}{\bibfnamefont{S.}~\bibnamefont{Moradi}},
  \bibinfo{journal}{Phys. Plasmas}
  \href{http://dx.doi.org/10.1063/1.3569841}{\textbf{\bibinfo{volume}{18}},
  \bibinfo{pages}{030703}} (\bibinfo{year}{2011}).

\bibitem[{\citenamefont{Angioni et~al.}(2012)\citenamefont{Angioni, Casson,
  Veth et~al.}}]{angioni_analytic_2012}
\bibinfo{author}{\bibfnamefont{C.}~\bibnamefont{Angioni}},
  \bibinfo{author}{\bibfnamefont{F.~J.} \bibnamefont{Casson}},
  \bibinfo{author}{\bibfnamefont{C.}~\bibnamefont{Veth}}, \bibnamefont{et~al.},
  \bibinfo{journal}{Phys. Plasmas}
  \href{http://dx.doi.org/doi:10.1063/1.4773051}{\textbf{\bibinfo{volume}{19}},
  \bibinfo{pages}{122311}} (\bibinfo{year}{2012}).

\bibitem[{\citenamefont{Moll{\'e}n et~al.}(2012)\citenamefont{Moll{\'e}n,
  Pusztai, F{\"u}l{\"o}p et~al.}}]{mollen_effect_2012}
\bibinfo{author}{\bibfnamefont{A.}~\bibnamefont{Moll{\'e}n}},
  \bibinfo{author}{\bibfnamefont{I.}~\bibnamefont{Pusztai}},
  \bibinfo{author}{\bibfnamefont{T.}~\bibnamefont{F{\"u}l{\"o}p}},
  \bibnamefont{et~al.}, \bibinfo{journal}{Phys. Plasmas}
  \href{http://dx.doi.org/10.1063/1.4719711}{\textbf{\bibinfo{volume}{19}},
  \bibinfo{pages}{052307}} (\bibinfo{year}{2012}).

\bibitem[{\citenamefont{Angioni and
  Helander}(2014)}]{angioni_neoclassical_2014}
\bibinfo{author}{\bibfnamefont{C.}~\bibnamefont{Angioni}} \bibnamefont{and}
  \bibinfo{author}{\bibfnamefont{P.}~\bibnamefont{Helander}},
  \bibinfo{journal}{{PPCF} In Press}  (\bibinfo{year}{2014}).

\bibitem[{\citenamefont{Belli and Candy}(2014)}]{belli_ps_2014}
\bibinfo{author}{\bibfnamefont{E.}~\bibnamefont{Belli}} \bibnamefont{and}
  \bibinfo{author}{\bibfnamefont{J.}~\bibnamefont{Candy}},
  \bibinfo{journal}{{PPCF} in press}  (\bibinfo{year}{2014}).

\bibitem[{\citenamefont{Moll{\'e}n et~al.}(2014)\citenamefont{Moll{\'e}n,
  Pusztai, Reinke et~al.}}]{mollen_impurity_2014}
\bibinfo{author}{\bibfnamefont{A.}~\bibnamefont{Moll{\'e}n}},
  \bibinfo{author}{\bibfnamefont{I.}~\bibnamefont{Pusztai}},
  \bibinfo{author}{\bibfnamefont{M.~L.} \bibnamefont{Reinke}},
  \bibnamefont{et~al.}, \bibinfo{journal}{{arXiv:1402.0309} [physics]}
  (\bibinfo{year}{2014}), \href{http://arxiv.org/abs/1402.0309}{[link]}.

\bibitem[{\citenamefont{Guirlet et~al.}(2009)\citenamefont{Guirlet, Villegas,
  Parisot et~al.}}]{guirlet_anomalous_2009}
\bibinfo{author}{\bibfnamefont{R.}~\bibnamefont{Guirlet}},
  \bibinfo{author}{\bibfnamefont{D.}~\bibnamefont{Villegas}},
  \bibinfo{author}{\bibfnamefont{T.}~\bibnamefont{Parisot}},
  \bibnamefont{et~al.}, \bibinfo{journal}{Nucl. Fusion}
  \href{http://dx.doi.org/10.1088/0029-5515/49/5/055007}{\textbf{\bibinfo{volu%
me}{49}}, \bibinfo{pages}{055007}} (\bibinfo{year}{2009}).

\bibitem[{\citenamefont{Angioni et~al.}(2011)\citenamefont{Angioni, McDermott,
  Fable et~al.}}]{angioni_gyrokinetic_2011}
\bibinfo{author}{\bibfnamefont{C.}~\bibnamefont{Angioni}},
  \bibinfo{author}{\bibfnamefont{R.}~\bibnamefont{McDermott}},
  \bibinfo{author}{\bibfnamefont{E.}~\bibnamefont{Fable}},
  \bibnamefont{et~al.}, \bibinfo{journal}{Nucl. Fusion}
  \href{http://dx.doi.org/10.1088/0029-5515/51/2/023006}{\textbf{\bibinfo{volu%
me}{51}}, \bibinfo{pages}{023006}} (\bibinfo{year}{2011}).

\bibitem[{\citenamefont{Howard et~al.}(2012)\citenamefont{Howard, Greenwald,
  Mikkelsen et~al.}}]{howard_quantitative_2012}
\bibinfo{author}{\bibfnamefont{N.}~\bibnamefont{Howard}},
  \bibinfo{author}{\bibfnamefont{M.}~\bibnamefont{Greenwald}},
  \bibinfo{author}{\bibfnamefont{D.}~\bibnamefont{Mikkelsen}},
  \bibnamefont{et~al.}, \bibinfo{journal}{Nucl. Fusion}
  \href{http://dx.doi.org/10.1088/0029-5515/52/6/063002}{\textbf{\bibinfo{volu%
me}{52}}, \bibinfo{pages}{063002}} (\bibinfo{year}{2012}).

\bibitem[{\citenamefont{Casson et~al.}(2013)\citenamefont{Casson, McDermott,
  Angioni et~al.}}]{casson_validation_2013}
\bibinfo{author}{\bibfnamefont{F.~J.} \bibnamefont{Casson}},
  \bibinfo{author}{\bibfnamefont{R.~M.} \bibnamefont{McDermott}},
  \bibinfo{author}{\bibfnamefont{C.}~\bibnamefont{Angioni}},
  \bibnamefont{et~al.}, \bibinfo{journal}{Nucl. Fusion}
  \href{http://dx.doi.org/10.1088/0029-5515/53/6/063026}{\textbf{\bibinfo{volu%
me}{53}}, \bibinfo{pages}{063026}} (\bibinfo{year}{2013}).

\bibitem[{\citenamefont{Henderson et~al.}(2014)}]{henderson_neoclassical_2014}
\bibinfo{author}{\bibfnamefont{S.}~\bibnamefont{Henderson}}
  \bibnamefont{et~al.}, \bibinfo{journal}{{PPCF} submitted}
  (\bibinfo{year}{2014}).

\bibitem[{\citenamefont{Angioni et~al.}(2014)}]{angioni_tungsten_2014}
\bibinfo{author}{\bibfnamefont{C.}~\bibnamefont{Angioni}} \bibnamefont{et~al.},
  \bibinfo{journal}{Nuclear Fusion (in Press)}  (\bibinfo{year}{2014}).

\bibitem[{\citenamefont{Peeters
  et~al.}(2009{\natexlab{a}})\citenamefont{Peeters, Camenen, Casson
  et~al.}}]{peeters_nonlinear_2009}
\bibinfo{author}{\bibfnamefont{A.~G.} \bibnamefont{Peeters}},
  \bibinfo{author}{\bibfnamefont{Y.}~\bibnamefont{Camenen}},
  \bibinfo{author}{\bibfnamefont{F.~J.} \bibnamefont{Casson}},
  \bibnamefont{et~al.}, \bibinfo{journal}{Computer Physics Communications}
  \href{http://dx.doi.org/10.1016/j.cpc.2009.07.001}{\textbf{\bibinfo{volume}{%
180}}, \bibinfo{pages}{2650}} (\bibinfo{year}{2009}{\natexlab{a}}).

\bibitem[{\citenamefont{Belli and Candy}(2008)}]{belli_kinetic_2008}
\bibinfo{author}{\bibfnamefont{E.~A.} \bibnamefont{Belli}} \bibnamefont{and}
  \bibinfo{author}{\bibfnamefont{J.}~\bibnamefont{Candy}},
  \bibinfo{journal}{Plasma Phys. Control. Fusion}
  \href{http://www.iop.org/EJ/abstract/0741-3335/50/9/095010}{\textbf{\bibinfo%
{volume}{50}}, \bibinfo{pages}{095010}} (\bibinfo{year}{2008}).

\bibitem[{\citenamefont{Belli and Candy}(2012)}]{belli_full_2012}
\bibinfo{author}{\bibfnamefont{E.~A.} \bibnamefont{Belli}} \bibnamefont{and}
  \bibinfo{author}{\bibfnamefont{J.}~\bibnamefont{Candy}},
  \bibinfo{journal}{Plasma Phys. Control. Fusion}
  \href{http://dx.doi.org/10.1088/0741-3335/54/1/015015}{\textbf{\bibinfo{volu%
me}{54}}, \bibinfo{pages}{015015}} (\bibinfo{year}{2012}).

\bibitem[{\citenamefont{Casson et~al.}(2010)\citenamefont{Casson, Peeters,
  Angioni et~al.}}]{casson_gyrokinetic_2010}
\bibinfo{author}{\bibfnamefont{F.~J.} \bibnamefont{Casson}},
  \bibinfo{author}{\bibfnamefont{A.~G.} \bibnamefont{Peeters}},
  \bibinfo{author}{\bibfnamefont{C.}~\bibnamefont{Angioni}},
  \bibnamefont{et~al.}, \bibinfo{journal}{Phys. Plasmas}
  \href{http://dx.doi.org/10.1063/1.3491110}{\textbf{\bibinfo{volume}{17}},
  \bibinfo{pages}{102305}} (\bibinfo{year}{2010}).

\bibitem[{\citenamefont{Belli and Candy}(2009)}]{belli_eulerian_2009}
\bibinfo{author}{\bibfnamefont{E.~A.} \bibnamefont{Belli}} \bibnamefont{and}
  \bibinfo{author}{\bibfnamefont{J.}~\bibnamefont{Candy}},
  \bibinfo{journal}{Plasma Phys. Control. Fusion}
  \href{http://dx.doi.org/10.1088/0741-3335/51/7/075018}{\textbf{\bibinfo{volu%
me}{51}}, \bibinfo{pages}{075018}} (\bibinfo{year}{2009}).

\bibitem[{\citenamefont{P{\"u}tterich et~al.}(2013)\citenamefont{P{\"u}tterich,
  Dux, Neu et~al.}}]{putterich_observations_2013}
\bibinfo{author}{\bibfnamefont{T.}~\bibnamefont{P{\"u}tterich}},
  \bibinfo{author}{\bibfnamefont{R.}~\bibnamefont{Dux}},
  \bibinfo{author}{\bibfnamefont{R.}~\bibnamefont{Neu}}, \bibnamefont{et~al.},
  \bibinfo{journal}{Plasma Phys. Control. Fusion}
  \href{http://dx.doi.org/10.1088/0741-3335/55/12/124036}{\textbf{\bibinfo{vol%
ume}{55}}, \bibinfo{pages}{124036}} (\bibinfo{year}{2013}).

\bibitem[{\citenamefont{Giroud et~al.}(2014)}]{giroud_this_2014}
\bibinfo{author}{\bibfnamefont{C.}~\bibnamefont{Giroud}} \bibnamefont{et~al.},
  \bibinfo{journal}{EPS Berlin Invited, PPCF}  (\bibinfo{year}{2014}).

\bibitem[{\citenamefont{Goniche et~al.}(2014)}]{goniche_this_2014}
\bibinfo{author}{\bibfnamefont{D.}~\bibnamefont{Goniche}} \bibnamefont{et~al.},
  \bibinfo{journal}{Proc. EPS Berlin}
  \href{http://ocs.ciemat.es/EPS2014PAP/pdf/O4.129.pdf}{
  \bibinfo{pages}{O4.129}} (\bibinfo{year}{2014}).

\bibitem[{\citenamefont{Puiatti et~al.}(2006)\citenamefont{Puiatti, Valisa,
  Angioni et~al.}}]{puiatti_analysis_2006}
\bibinfo{author}{\bibfnamefont{M.~E.} \bibnamefont{Puiatti}},
  \bibinfo{author}{\bibfnamefont{M.}~\bibnamefont{Valisa}},
  \bibinfo{author}{\bibfnamefont{C.}~\bibnamefont{Angioni}},
  \bibnamefont{et~al.}, \bibinfo{journal}{Phys. Plasmas}
  \href{http://dx.doi.org/10.1063/1.2187424}{\textbf{\bibinfo{volume}{13}},
  \bibinfo{pages}{042501}} (\bibinfo{year}{2006}).

\bibitem[{\citenamefont{Casson}(2011)}]{casson_turbulent_2011}
\bibinfo{author}{\bibfnamefont{F.~J.} \bibnamefont{Casson}},
  \href{http://wrap.warwick.ac.uk/36765/}{\bibinfo{type}{{Ph.D.}}},
  \bibinfo{school}{University of Warwick} (\bibinfo{year}{2011}).

\bibitem[{\citenamefont{Mantica et~al.}(2014)}]{mantica_this_2014}
\bibinfo{author}{\bibfnamefont{P.}~\bibnamefont{Mantica}} \bibnamefont{et~al.},
  \bibinfo{journal}{Proc. EPS Berlin}
  \href{http://ocs.ciemat.es/EPS2014PAP/pdf/P1.017.pdf}{
  \bibinfo{pages}{P1.017}} (\bibinfo{year}{2014}).

\bibitem[{\citenamefont{Peeters
  et~al.}(2009{\natexlab{b}})\citenamefont{Peeters, Strintzi, Camenen
  et~al.}}]{peeters_influence_2009}
\bibinfo{author}{\bibfnamefont{A.~G.} \bibnamefont{Peeters}},
  \bibinfo{author}{\bibfnamefont{D.}~\bibnamefont{Strintzi}},
  \bibinfo{author}{\bibfnamefont{Y.}~\bibnamefont{Camenen}},
  \bibnamefont{et~al.}, \bibinfo{journal}{Phys. Plasmas}
  \href{http://dx.doi.org/10.1063/1.3097263}{\textbf{\bibinfo{volume}{16}},
  \bibinfo{pages}{042310}} (\bibinfo{year}{2009}{\natexlab{b}}).

\bibitem[{\citenamefont{Camenen et~al.}(2009)\citenamefont{Camenen, Peeters,
  Angioni et~al.}}]{camenen_impact_2009}
\bibinfo{author}{\bibfnamefont{Y.}~\bibnamefont{Camenen}},
  \bibinfo{author}{\bibfnamefont{A.~G.} \bibnamefont{Peeters}},
  \bibinfo{author}{\bibfnamefont{C.}~\bibnamefont{Angioni}},
  \bibnamefont{et~al.}, \bibinfo{journal}{Phys. Plasmas}
  \href{http://dx.doi.org/10.1063/1.3057356}{\textbf{\bibinfo{volume}{16}},
  \bibinfo{pages}{012503}} (\bibinfo{year}{2009}).

\bibitem[{\citenamefont{Cenacchi and Taroni}(1988)}]{cenacchi_jetto:_1988}
\bibinfo{author}{\bibfnamefont{G.}~\bibnamefont{Cenacchi}} \bibnamefont{and}
  \bibinfo{author}{\bibfnamefont{A.}~\bibnamefont{Taroni}},
  \bibinfo{journal}{{JET} Internal report}  \bibinfo{pages}{{JET--IR(88)03}}
  (\bibinfo{year}{1988}).

\bibitem[{\citenamefont{Romanelli et~al.}(2014)\citenamefont{Romanelli, Parail,
  Corrigan et~al.}}]{romanelli_jintrac:_2014}
\bibinfo{author}{\bibfnamefont{M.}~\bibnamefont{Romanelli}},
  \bibinfo{author}{\bibfnamefont{V.}~\bibnamefont{Parail}},
  \bibinfo{author}{\bibfnamefont{G.}~\bibnamefont{Corrigan}},
  \bibnamefont{et~al.}, \bibinfo{journal}{Plasma and Fusion research}
  \href{http://www.jspf.or.jp/PFR/PFR_articles/pfr2014S2/pfr2014_09-3403023.ht%
ml}{\textbf{\bibinfo{volume}{9}}, \bibinfo{pages}{3403023}}
  (\bibinfo{year}{2014}).

\bibitem[{\citenamefont{Fable et~al.}(2012)\citenamefont{Fable, Angioni,
  Fischer et~al.}}]{fable_progress_2012}
\bibinfo{author}{\bibfnamefont{E.}~\bibnamefont{Fable}},
  \bibinfo{author}{\bibfnamefont{C.}~\bibnamefont{Angioni}},
  \bibinfo{author}{\bibfnamefont{R.}~\bibnamefont{Fischer}},
  \bibnamefont{et~al.}, \bibinfo{journal}{Nucl. Fusion}
  \href{http://dx.doi.org/10.1088/0029-5515/52/6/063017}{\textbf{\bibinfo{volu%
me}{52}}, \bibinfo{pages}{063017}} (\bibinfo{year}{2012}).

\bibitem[{\citenamefont{Hobirk et~al.}(2012)\citenamefont{Hobirk, Schweinzer,
  Orte et~al.}}]{hobirk_overview_2012}
\bibinfo{author}{\bibfnamefont{J.}~\bibnamefont{Hobirk}},
  \bibinfo{author}{\bibfnamefont{J.}~\bibnamefont{Schweinzer}},
  \bibinfo{author}{\bibfnamefont{L.~B.} \bibnamefont{Orte}},
  \bibnamefont{et~al.}, \bibinfo{journal}{{IAEA} {FEC}, San Diego}
  \href{http://www-naweb.iaea.org/napc/physics/FEC/FEC2012/papers/401_EXP203.p%
df}{ \bibinfo{pages}{EX/P2--03}} (\bibinfo{year}{2012}).

\bibitem[{\citenamefont{Odstrcil et~al.}(2012)\citenamefont{Odstrcil, Mlynar,
  Odstrcil et~al.}}]{odstrcil_modern_2012}
\bibinfo{author}{\bibfnamefont{M.}~\bibnamefont{Odstrcil}},
  \bibinfo{author}{\bibfnamefont{J.}~\bibnamefont{Mlynar}},
  \bibinfo{author}{\bibfnamefont{T.}~\bibnamefont{Odstrcil}},
  \bibnamefont{et~al.}, \bibinfo{journal}{Nucl. Inst. Meth. Phys. Res. Sec. A}
  \href{http://dx.doi.org/10.1016/j.nima.2012.05.063}{\textbf{\bibinfo{volume}%
{686}}, \bibinfo{pages}{156}} (\bibinfo{year}{2012}).

\bibitem[{\citenamefont{Brambilla}(1999)}]{brambilla_numerical_1999}
\bibinfo{author}{\bibfnamefont{M.}~\bibnamefont{Brambilla}},
  \bibinfo{journal}{Plasma Phys. Control. Fusion}
  \href{http://dx.doi.org/10.1088/0741-3335/41/1/002}{\textbf{\bibinfo{volume}%
{41}}, \bibinfo{pages}{1}} (\bibinfo{year}{1999}).

\bibitem[{\citenamefont{Bilato et~al.}(2011)\citenamefont{Bilato, Brambilla,
  Maj et~al.}}]{bilato_simulations_2011}
\bibinfo{author}{\bibfnamefont{R.}~\bibnamefont{Bilato}},
  \bibinfo{author}{\bibfnamefont{M.}~\bibnamefont{Brambilla}},
  \bibinfo{author}{\bibfnamefont{O.}~\bibnamefont{Maj}}, \bibnamefont{et~al.},
  \bibinfo{journal}{Nucl. Fusion}
  \href{http://dx.doi.org/10.1088/0029-5515/51/10/103034}{\textbf{\bibinfo{vol%
ume}{51}}, \bibinfo{pages}{103034}} (\bibinfo{year}{2011}).

\bibitem[{\citenamefont{Brambilla}(1994)}]{brambilla_quasi-linear_1994}
\bibinfo{author}{\bibfnamefont{M.}~\bibnamefont{Brambilla}},
  \bibinfo{journal}{Nucl. Fusion}
  \href{http://dx.doi.org/10.1088/0029-5515/34/8/I06}{\textbf{\bibinfo{volume}%
{34}}, \bibinfo{pages}{1121}} (\bibinfo{year}{1994}).

\bibitem[{\citenamefont{Angioni et~al.}(2009)\citenamefont{Angioni, Fable,
  Greenwald et~al.}}]{angioni_particle_2009}
\bibinfo{author}{\bibfnamefont{C.}~\bibnamefont{Angioni}},
  \bibinfo{author}{\bibfnamefont{E.}~\bibnamefont{Fable}},
  \bibinfo{author}{\bibfnamefont{M.}~\bibnamefont{Greenwald}},
  \bibnamefont{et~al.}, \bibinfo{journal}{Plasma Phys. Control. Fusion}
  \href{http://dx.doi.org/10.1088/0741-3335/51/12/124017}{\textbf{\bibinfo{vol%
ume}{51}}, \bibinfo{pages}{124017}} (\bibinfo{year}{2009}).

\bibitem[{\citenamefont{P{\"u}tterich et~al.}(2012)}]{putterich_tungsten_2012}
\bibinfo{author}{\bibfnamefont{T.}~\bibnamefont{P{\"u}tterich}}
  \bibnamefont{et~al.}, \bibinfo{journal}{{IAEA} {FEC}, San Diego}
  \href{http://www.iop.org/Jet/fulltext/EFDC120622.pdf}{
  \bibinfo{pages}{EX/P3--15}} (\bibinfo{year}{2012}).

\end{thebibliography}

\end{document}